\documentclass[smallextended,authoryear]{svjour3}       
\smartqed  
 \usepackage{mathptmx}      
\usepackage{amssymb}
\usepackage{color}
\usepackage[cp1250]{inputenc}
\usepackage{xspace}
\usepackage[round]{natbib}
\usepackage{multibib}

\usepackage{amsfonts}
\usepackage{amsmath}
\usepackage[english]{babel}
\usepackage{verbatim}

\usepackage{amsthm}

\ifx\pdfoutput\undefined
\usepackage{graphicx}
\else
\usepackage[pdftex]{graphicx}
\usepackage{epstopdf}
\fi

\newcommand{\network}{\mathcal{N}}
\newcommand{\nodes}{\mathcal{V}}
\newcommand{\links}{\mathcal{L}}
\newcommand{\W}{\mathbf{W}}
\newcommand{\A}{\mathbf{A}}
\newcommand{\J}{\mathcal{J}}
\newcommand{\K}{\mathcal{K}}
\newcommand{\M}{\mathcal{M}}
\newcommand{\U}{\mathcal{U}}
\newcommand{\n}[1]{\mathcal{#1}}
\newcommand{\outdeg}{\mathrm{outdeg}}
\newcommand{\diag}{\mathrm{diag}}
\begin{document}

\title{Network analysis of Zentralblatt MATH data}


\author{Monika Cerin\v sek \and Vladimir Batagelj}


\institute{V. Batagelj \at
               University of Ljubljana, FMF, Department of Mathematics,
   Jadranska 19, 1000 Ljubljana, Slovenia\\
              \email{vladimir.batagelj@fmf.uni-lj.si}                        \\
              URL: http://pajek.imfm.si
           \and
           M. Cerin\v sek \at
              Hru\v ska d.o.o., Kajuhova 90, 1000 Ljubljana\\
              Tel.: +386-31421590\\
              Fax: +386-59022240\\
              \email{monika@hruska.si}
}

\date{}

\maketitle

\begin{abstract}
We analyze the data about works (papers, books) from the time period 1990-2010 that are collected in Zentralblatt MATH database. The data were converted into four 2-mode networks (works $\times$ authors, works $\times$ journals, works $\times$ keywords and works $\times$ MSCs) and into a partition of works by publication year. The networks were analyzed using Pajek -- a program for analysis and visualization of large networks. We explore the distributions of some properties of works and the collaborations among mathematicians. We also take a closer look at the characteristics of the field of graph theory as were realized with the publications.
\end{abstract}

%
%

\section{Introduction}
Bibliographic data allow us to explore the development of an area of research, which authors collaborated most, in which areas of research exist stronger collaboration groups, in which areas authors prefer to work alone or in smaller groups, and much more. 
Analysis of bibliographic data does not contribute to the areas of research directly, but helps us to understand how they are structured. 
Network analysis of bibliographic data has been already widely explored, started with E.~Garfield \citep{garfield} on. 
In the paper we intend to present an insight into the field of mathematics as recorded by the Zentralblatt MATH (ZB) database in the decades 1990-2010.  
The ZB database is maintained by the Berlin editorial office of FIZ Karlsruhe in cooperation with European academies and mathematical institutes.

In cooperation with prof.~Bernd Wegner and his associates at FIZ Karlsruhe we obtained in January 2011 the basic data about works (papers, books) for the time period 1990-2010 that are collected in the ZB database. We chose to explore this bibliographic data using network analysis. For computations we used the program Pajek \citep{pajek, pajek1}, a tool for analysis and visualization of large networks. 
In this paper we present the results from basic network analyses (statistical information about the data) and identification of important elements (authors, keywords, and journals).

In the paper we first describe the data and discuss some problems encountered in transforming the data into networks. In the third section, different distributions are presented. The analysis of the collaboration network among mathematicians is presented in the fourth section. In the last section we take a closer look on the selected area of mathematics -- the graph theory. Analysis of the collaborations among graph theorists,  graph theory determining keywords, journals biased toward graph theory and areas of mathematics that overlap with graph theory are presented.

%
%
\section{Data}
The data obtained from the ZB database contain several information about each work. The collection of the information about a single work is called a record and is composed of different fields. Each field has its own $2$-character identifier:
\begin{itemize}
\item[] an -- identification number of a work (set by ZB),
\item[] ai -- unified author's name,
\item[] au -- author's name,
\item[] py -- publication year,
\item[] cc -- classification (Mathematical Subject Classification - MSC) \citep{mscji},
\item[] ti -- title,
\item[] ut -- keywords,
\item[] is -- journal's International Standard Serial Number (ISSN),
\item[] so -- journal's title, pages, year,
\item[] se -- data about journal (identification number by ZB, whole and short title, ISSN).
\end{itemize}
An example of a record:
{\small
\begin{verbatim}
an  01714102
ai  -; sastre-vazquez.patricia; -
is  ISSN 0368-492X
au  Us\'o-Dom\`enech, J.L.; Sastre-Vazquez, P.; Mateu, J.
py  2001
cc  *68U20
ti  Syntax and first entropic approximation of $L(M_T)$. A 
      language for ecological modelling.
ut  modelling process; text-model based language
so  Kybernetes 30, No.9-10, 1304-1317 (2001).
se  00000540	Kybernetes	Kybernetes	0368-492X
\end{verbatim}
}

\subsection{Problems with the data}
Data about works are entered in the ZB database by editors. The most common problem are the missing data. Some papers and books do not have all types of information entered. If the missing information is needed in any of analyses this leeds to additional problems.

The non ASCII characters in the text are represented by \TeX{} commands.
The problem is the nonuniform use of \TeX. For example

{\small
\begin{verbatim}
au  Must\u{a}\c{t}a, Costic\u{a}
au  Must\u a\c ta, Costic\u a
\end{verbatim}
}
\noindent
are two different writings of the name of the same author. To solve the problem we need to write a script that  recognizes all the different writings of the same character.

Author's names are only partially unified in the ZB database. Some authors have unified names, but others do not, or even have several of them (synonymy). This problem is not easily solvable, because someone would need to look at the list of all authors and their unified names and make the necessary corrections. There exist authors with their names written in more than one variant. For example Manko\v c Bor\v stnik, Norma Susana is written as

{\small
\begin{verbatim}
Bor\v stnik, N. S. Manko\v c
Manko\v c Bor\v stnik, N.
Manko\v c-Bor\v stnik, Norma
Manko\v c Bor\v stnik, Norma Susana
Mankoc-Borstnik, N.S.
Manko\v c Bor\v stnik, N.S.
\end{verbatim}
}

The unification is also a problem for another reason. Some authors have very similar or even the same names (homonymy). They might also have the same unification of their name in ZB database, which results in the problem of distinguishing between these authors. 

Not only the authors' names are the problem, but also the keywords.
Because not all of the works have assigned keywords, we extracted also words from the title and considered them as keywords.
(Real) keywords are actually phrases consisting of at least one word. We splitted phrases into words and removed the stop words. Related keywords were unified using lemmatization (MontyLingua package in Python). For example, keywords algebra and algebras were unified.

Journals in ZB have identification numbers. This, in principle, solves the unique identification problem. But we did find one journal with two identifiers during analyses:

{\small
\begin{verbatim}
se  00000552 Match Match 0340-6253
se  00003047 MATCH - Communications in Mathematical and in 
             Computer Chemistry MATCH Commun. Math. Comput. 
             Chem. 0340-6253
\end{verbatim}
}
\noindent
We treat them as a single journal.

Journals are changing through time -- a new journal is `born', a journal `dies', some journals are merged into journal, a journal is split into some journals, a journal changes the title, etc. Some journals had just changed the title. Because of such changes they appear as different journals in the database. We merged different appearances of a journal that changed the title into one journal.

\subsection{Preparation of the data}
With a special program written in Python we converted the data into the Pajek format \citep{pajek}. We obtained four compatible $2$-mode networks and a partition of works by their publication year.

A network is a structure $\network = (\nodes, \links, w)$, which consists of a set of nodes $\nodes$, a set of links among nodes $\links$ and a weight function $w: \links \rightarrow \mathbb{R}$, that determines the weights of the links. A network is called a $2$-mode network, if the set of nodes $\nodes$ is partitioned into two disjoint subsets and each link has its end nodes in different subsets.

In our data, the first subset of nodes in all four networks consists of identifiers of \textbf{works} and is denoted by $\mathcal{W}$. Nodes in the second subset represent one of the following:
\begin{itemize}
\item $\mathcal{A}$ -- a set of \textbf{authors},
\item $\J$ -- a set of \textbf{journals},
\item $\K$ -- a set of \textbf{keywords},
\item $\M$ -- a set of \textbf{MSCs} (mathematical subject classifications).
\end{itemize}

Information was extracted from the records of all works. The identifiers of works were extracted from the field \verb$an$, journals from the field \verb$se$ and MSCs from the field \verb$cc$. Keywords were extracted from the fields \verb$ut$ and \verb$ti$ as phrases and then decomposed into words and unified using lemmatization. Names of authors were extracted from the field \verb$ai$. If the author's ZB-unified name does not exist, his/her name was extracted from the field \verb$au$ and unified into ZB-names-like form.

We expect that most of the important mathematicians have their unified name. For the rest, we decided to treat the synonymy/homonymy as a kind of noise and reconsider them in cases when they appear as `duplicates' in the results.

Links in all produced networks are directed -- \textbf{arcs}, and they link each work to some representatives in the second set. The co-authorship network of works $\times$ authors $\W\A = ((\mathcal{W}, \mathcal{A}), \links, w)$ is a network in which each work is linked to all of its authors, $(p,i) \in \links \Leftrightarrow$ $i$ is an author of work $p$. The other three networks are defined in a similar way -- works are linked to journals, keywords and MSCs in networks works $\times$ journals $\W\mathbf{J}$, works $\times$ keywords $\W\mathbf{K}$, and works $\times$ classifications $\W\mathbf{M}$, respectively. We will also use a simplified notation for a transposed network: the transposed network of the network $\W\A$ is denoted with $\A\W \equiv \W\A^T$ and is obtained from $\W\A$ by changing the directions of all its arcs. The sizes of all four networks are listed in Table~\ref{sizes}.

\begin{table}[!h]
\centering
\caption{\small{Sizes of $2$-mode networks.}
}
\vspace{5pt}
\begin{tabular}{r||r|r|r|r}
Network & $\W\A$ & $\W\mathbf{J}$ & $\W\mathbf{K}$ & $\W\mathbf{M}$ \\
\hline
\hline
Size of the first set & $1,339,201$ & $1,339,201$ & $1,339,201$ & $1,339,201$\\
Size of the second set & $557,104$ & $3,158$ & $143,513$ & $12,390$\\
Number of arcs & $2,550,437$ & $1,331,036$ & $15,062,377$ & $3,370,820$
\end{tabular}
\label{sizes}
\end{table}

As mentioned before, we had problems with the notion of an author. Some of them appeared twice or even more times under different names in the network $\W\A$ because of only the partial unification of their names. We made a partition of the set of authors by collecting different appearances of the same author. For example O'Regan, Donal is once written as \verb$oregan.donal$ and another time as \verb$o'regan.d$. This author has a ZB-unified name oregan.donal, but sometimes his unified name is not written and in such cases our program for the data conversion creates it from the full author's name -- O'Regan, Donal get unified-like name \verb$o'regan.d$. Another author with similar problem is Pe\v cari\' c, Josip E. His unified ZB-name is \verb$pecaric.josip-e$, sometimes unified name is not written and we get \verb$pecaric.j$ and \verb$pecaric.j-e$ because of two different writings of his full name: Pe\v{c}ari\'c, J. and Pe\v{c}ari\'c, J. E.  Yet another source of problems is the writing of Eastern European surnames: Krachkovskij, A. P., and Krachkovskii, A. P., are probably representing the same author.

The partition of author's names solved the unification problem only partially. We also used the AMS identification of authors~\citep{ams_id} for help with the unification problem. All the following analyses were made after the additional unification of different appearances of the same author names.

We also solved the problem with journals. Different names of the same journal were replaced by a single name -- from $3158$ journal names we obtained $2665$ unique journal names.

%
%
\section{Distributions of properties of works}\label{distributions}
We examined degrees of nodes in the obtained networks to determine distributions of different data. With outdegrees of nodes in the set of works in networks $\W\A,$ $\W\mathbf{K},$ $\W\mathbf{M}$ we obtained the distributions of works by number of authors, keywords and classifications. Each work is supposed to be published in at most one journal. None of the works in the database was published in more than one journal. There are $8165$ works that have no journal determined.

\begin{figure}[!h]
\centering
\includegraphics[width=\textwidth]{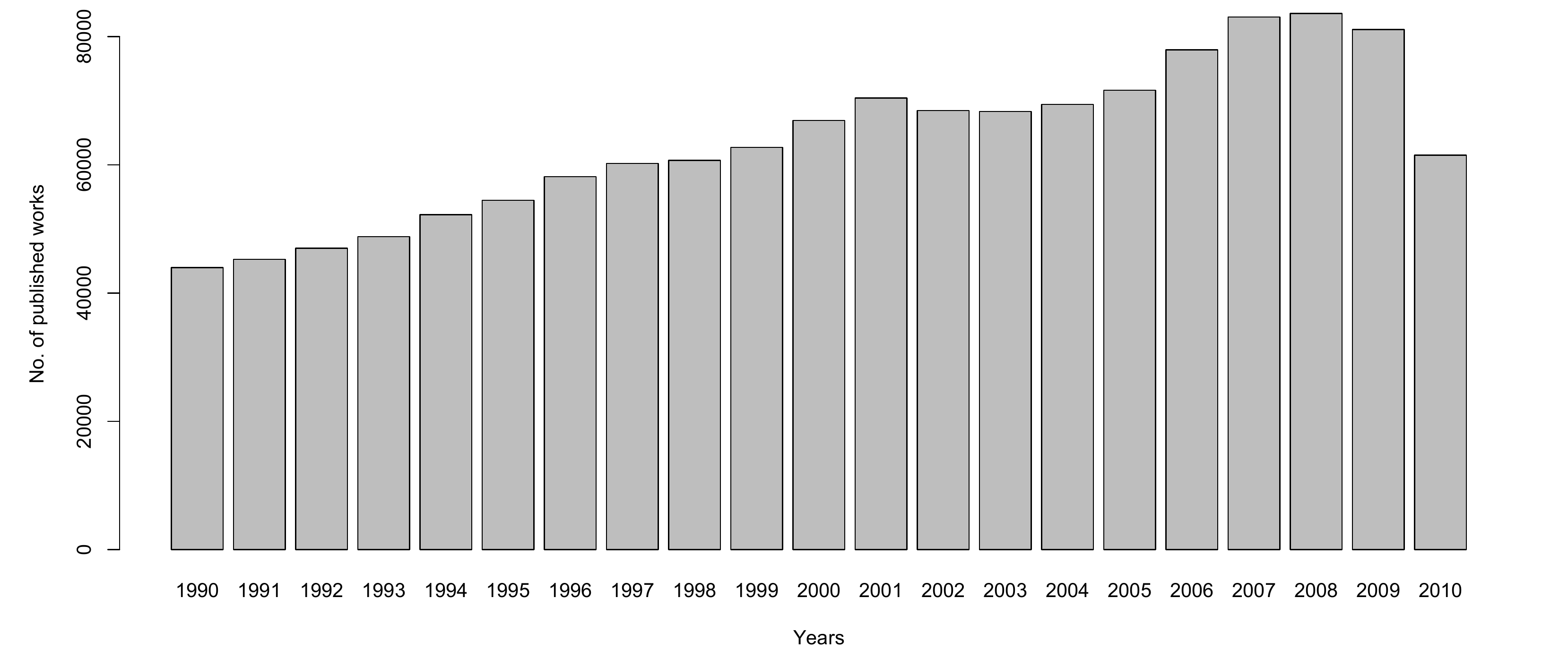}
\caption{\small{Distribution of works by their publication year}}
\label{years}
\end{figure}

The distribution of works published in the time period 1990-2010 that are indexed in Zentralblatt MATH by their publication year is shown in Fig.~\ref{years}. We see that the number of indexed works is growing -- in 20 years it has almost doubled. The decrease in the years 2009 and 2010 is due to works that are still to be indexed.

\begin{figure}[p]
\centering
\includegraphics[height=\textheight]{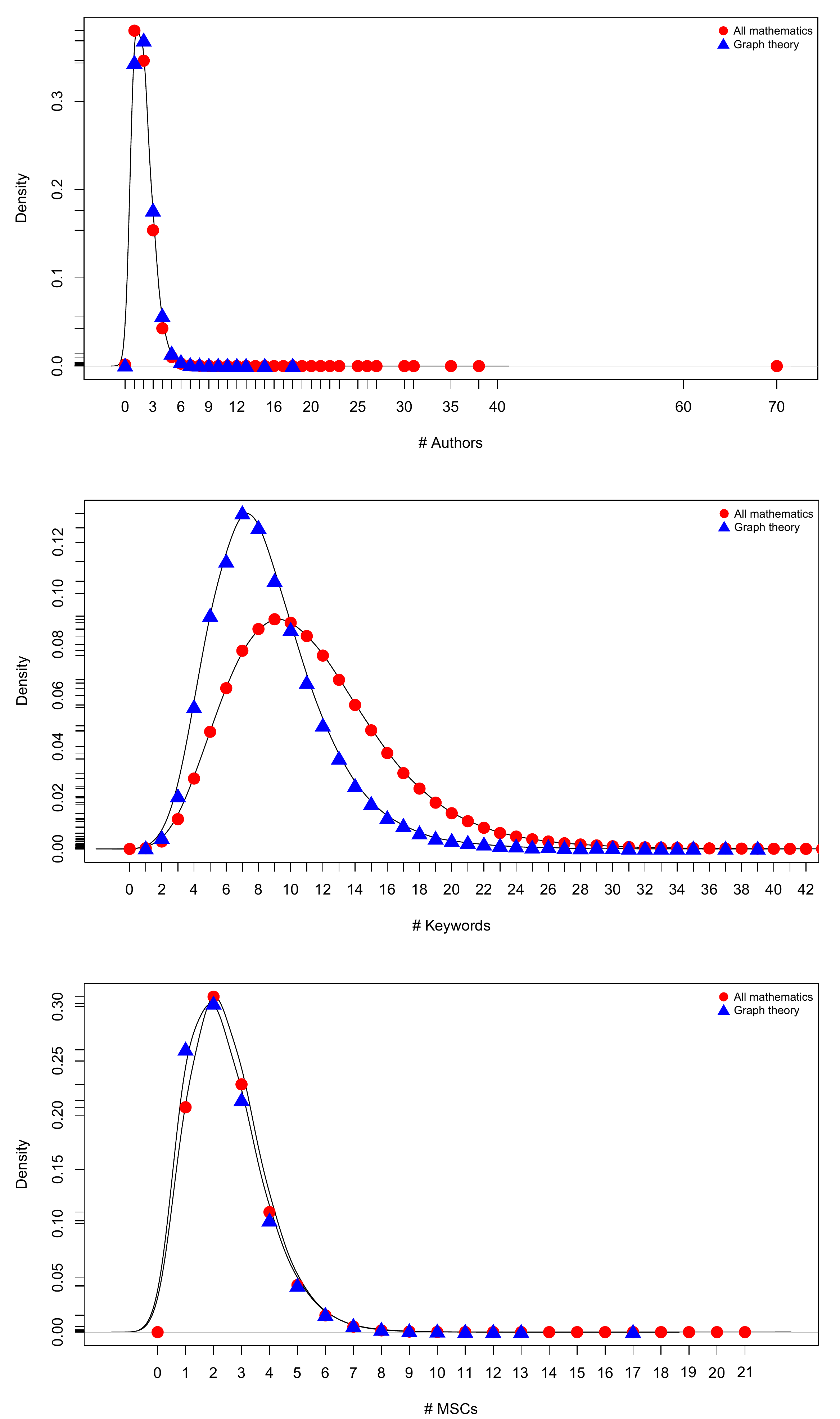}
\caption{\small{Distributions of works by the number of authors, keywords and MSC classifications. These figures include also the same distributions for only the field of graph theory which is the topic of Section \ref{05cxx}.}}
\label{density_gt}
\end{figure}

For a work $p \in \mathcal{W}$ its $\outdeg(p)$ in the network $\W\A$ is a number of authors of the work $p$.
Distribution of works by a number of authors, $\mathrm{F}(d) = |\{ p \in \mathcal{W}: \outdeg (p) = d \}| = $ number of works each having exactly $d$ authors, is shown in Fig.~\ref{density_gt} in the top diagram. The curves in all diagrams in Fig.~\ref{density_gt} are gaussian kernel density estimates of the distributions.
 More than $\frac{1}{3}$ of all works ($37.99\%$) was written by a single author and another $\frac{1}{3}$ of all works ($34.60\%$) by a pair of authors. $2383$ works do not have any author attributed and on the other hand some works have large number of co-authors -- even $70$ co-authors per work. Works with the largest number of co-authors are:
\begin{itemize}
\item $70$ co-authors --  Aderholz, M. et al.: Distributed applications monitoring at system and network level. Comput. Phys. Commun. 140, No.1-2, 219-225 (2001).
\item $38$ co-authors -- Bridle, S. et al.: Handbook for the GREAT08 challenge: an image analysis competition for cosmological lensing. Ann. Appl. Stat. 3, No. 1, 6-37 (2009).
\item $35$ co-authors -- Regan, S.P. et al.: Direct-drive inertial confinement fusion implosions on omega. Astrophys. Space Sci. 298, No. 1-2, 227-233 (2005).
\end{itemize}

The distribution of works by a number of keywords is shown in the center of Fig.~\ref{density_gt}. Note that our keywords were produced from the keywords and the title, as explained earlier. This distribution is quite flat. 
 Approximately $50 \%$ of all works have the number of keywords between $10$ and $18$. Works with the largest number of keywords are:
\begin{itemize}
\item $71$ keywords -- Baianu, I.C. et al.: Algebraic topology foundations of supersymmetry and symmetry breaking in quantum field theory and quantum gravity: a review. SIGMA, Symmetry Integrability Geom. Methods Appl. 5, Paper 051, 70 p., electronic only (2009).
\item $69$ keywords -- Dutta, H.: On some sequence spaces generated by  $\Delta (r)$ -- and  $\Delta r$ -- difference of infinite matrices. Int. J. Open Probl. Comput. Sci. Math., IJOPCM 2, No. 4, 496-504 (2009).
\item $68$ keywords -- Cheng B. and Tong, H.: On consistent nonparametric order determination and chaos. J. R. Stat. Soc., Ser. B 54, No.2, 427-449 (1992).
\end{itemize}

Distribution of works by number of MSCs is shown at the bottom of Fig.~\ref{density_gt}. Approximately one third of all works ($30.92\%$) were classified with two MSCs and approximately $40\%$ of all works ($43.57\%$) were classified with one or three MSCs. Works with largest numbers of MSCs are:
\begin{itemize}
\item $21$ MSCs -- Auroux, D. et al: Report 35/2006: Four-dimensional Manifolds (August 6th -- August 12th, 2006). Oberwolfach Rep. 3, No. 3, 2059-2140 (2006).
\item $21$ MSCs -- Dechevsky, L.T.: Concluding remarks to paper ``properties of function spaces generated by the averaged moduli of smoothness". Int. J. Pure Appl. Math. 49, No. 1, 147-152 (2008).
\item $20$ MSCs -- Aubin, J.-P.: A survey of viability theory. SIAM J. Control Optimization 28, No.4, 749-788 (1990).
\end{itemize}

It turns out that all distributions in Fig.~\ref{density_gt} can be very well approximated by the 
lognormal distribution, gamma distribution and also by the generalized reciprocal 
power exponential curve $c*(x+d)^\frac{a}{b+x}$. In all cases we get the best fit
with gamma distribution. The results are given in Table \ref{gamma}. For technical details see
Subsection~2.5.2 \textit{Fitting distributions} in \cite{tempo}.

\begin{table}[h]
\begin{center}
\caption{Fitting the gamma distribution $c \cdot \Gamma(x, a, b)$.}\label{gamma}

\begin{tabular}{l||rrrr}
 Distribution                     &          $c$        &   $a$    &   $b$   &  residual SS  \\
 \hline
\hline
Authors (all) & 			$1.342 \cdot 10^6$  &   $3.621$  &   $1.905$ &  $99893627$  \\
Authors (graph theory)     &	$4.056 \cdot 10^4$     &$3.876$  &   $1.920$ &     $23743$  \\
Keywords (all)    &       		$1.333 \cdot 10^6$  &   $5.561$  &   $0.498$&  $10079394$  \\
Keywords (graph theory)    &$4.006 \cdot 10^4$  &   $6.960$  &   $0.832$&  $188488$ \\
MSCs (all)            &       		$1.345 \cdot 10^6$  &   $3.718$  &   $1.457$ & $273424172$  \\
MSCs (graph theory)       &       $4.136 \cdot 10^4$     & $3.078$  &   $1.261$ &    $357960$  \\
\end{tabular}
\end{center}
\end{table}

\begin{figure}[p]
\centering
\includegraphics[height=\textheight]{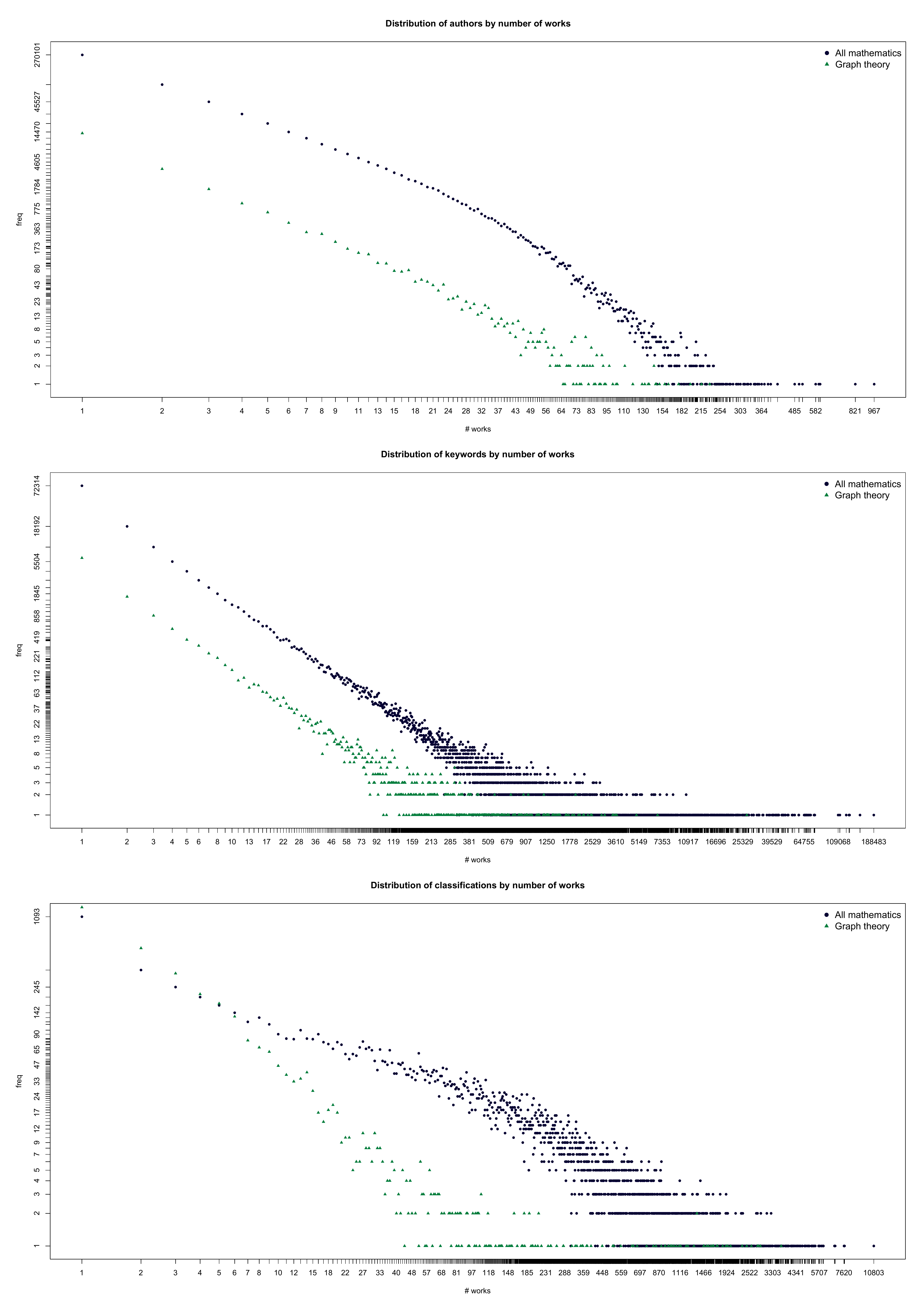}
\caption{\small{From top to bottom: Frequency distribution of authors by the number of co-authored works in a double logarithmic scale; Frequency distribution of keywords by the number of works using a keyword in their description in a double logarithmic scale; Frequency distribution of MSCs by the number of classified works in a double logarithmic scale. These figures also include the same distributions for only the field of graph theory which is the topic of Section \ref{05cxx}}}
\label{dist}
\end{figure}

In addition to examining the distributions of degrees of nodes in the first subset of the two-mode networks, we examined the distributions of degrees of nodes in the second subset as well.
The distribution of authors by number of works they co-authored is shown in Fig.~\ref{dist} in the top figure. For example a dot in the upper left corner represents $271013$ authors that each co-authored only one work in the time-period of 1990-2010. Dots in the lower right corner are representing Ballico, Edoardo with $967$ works co-authored in a given time-period, O'Regan, Donal with $821$ works, Pe\v{c}ari\'c, Josip with $606$ works, Agarwal, Ravi P. with $598$ works and Srivastava, H.M. with $582$ works co-authored in a given time-period as indexed in the ZB database. One can notice that Lotka's law holds for authors of up to $16$ works.

\begin{figure}
\includegraphics[width=\textwidth]{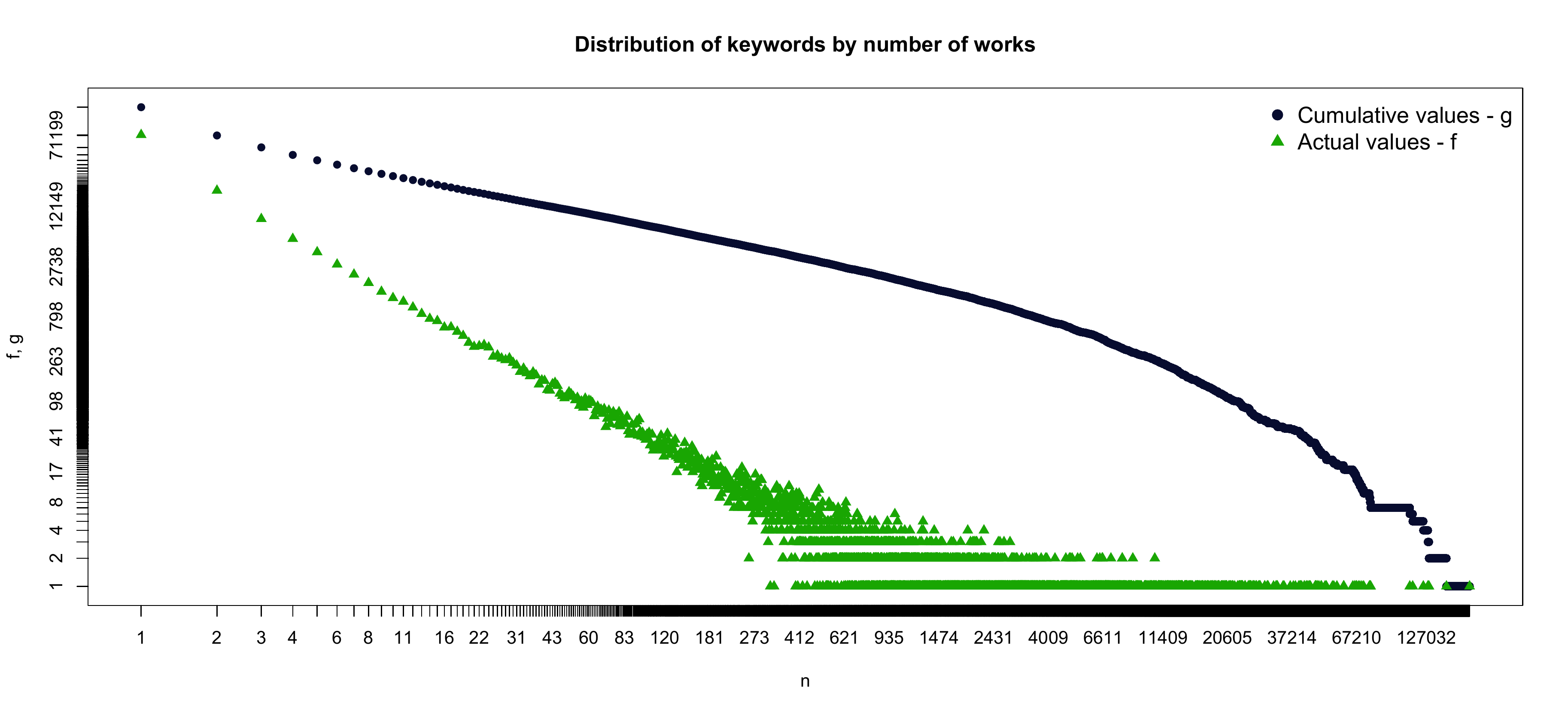}
\caption{Sequences $f$ and $g$ for the distribution of keywords by the number of works.
\label{power}}
\end{figure}

The distribution of keywords by the number of works they describe is shown in Fig.~\ref{dist} in the second figure. The dot in the upper left corner represents $72314$ keywords that were used in the description of works only once in the time-period 1990-2010. Dots in lower right corner represent most commonly used keywords: equation ($188483$ times used), problem ($152514$), function ($129957$), method ($128740$), model ($123448$), space ($112000$), solution ($109068$), linear ($76241$), theory ($75873$) and finite ($75398$). These words are actually the most common words in mathematics. The shape of the distribution of keywords by the number of works in in the second figure in Fig.~\ref{dist} is typical for empirical distributions of quantities following the power law $f_n = cn^{-\alpha}$.
Using the function power.law.fit in the R package igraph that implements M. Newman's
procedure described in \cite{powerlaw} we get $\alpha = 1.85$. 
To visually check the power law nature of the distribution we can use the property 
that, for $\alpha>1$, if the sequence $(f_n)$ obeys the power law then it is also obeyed by the sequence $(g_n)$ defined as $ g_n = \sum_{i=n}^\infty f_i \simeq C n^{1-\alpha} $
 as is presented with Eq. 4.38 in Barab\'{a}si, A-L.: \emph{Network Science}, 2014, available at http://barabasilab.com/networksciencebook.
Therefore in the joint picture of both sequences in double logarithmic scale we should
get two `lines'. For the distribution of keywords by the number of works this is not the
case as can be seen on Figure~\ref{power}. The distribution doesn't obey the power law.

The distribution of MSCs by the number of works that were classified with a given MSC is displayed in Fig.~\ref{dist} in the bottom figure. Every work is classified with one primary and maybe some secondary MSCs. The same primary and secondary MSC (for example 74S05 and $\star$74S05) are represented with one dot. Each dot represents on the y-axis determined number of MSCs that were used for classification of on x-axis determined number of works. The dot in the upper left corner represents $1093$ MSCs, that were included in the classification of only one work in the time-period of 1990-2010. MSCs in Table~\ref{tab_msc} are the most frequently used MSCs. These MSCs are represented with dots in the lower right corner of Fig.~\ref{dist}. The most frequently used primary MSCs are listed in Table~\ref{tab_msc1}.

\begin{table}[!h]
\centering
\caption{\small{Table of the most popular MSCs.}}
\vspace{5pt}
\scriptsize{
\begin{tabular}{@{}p{0.7cm}@{\hspace{3mm}}p{5.6cm}@{\hspace{3mm}}p{4.2cm}@{\hspace{3mm}}r@{}}
\textbf{MSC} & & & \textbf{No. of}\\
\textbf{code} & \textbf{$2$-char MSC name} & \textbf{MSC name} & \textbf{works}\\
\hline
80A20 & Classical thermodynamics, heat transfer & Heat and mass transfer, heat flow & $13279$\\ 
74S05 & Mechanics of deformable solids & Finite element methods & $13271$\\ 
68T05 & Computer science & Learning and adaptive systems & $11775$\\ 
35B40 & Biology and other natural sciences & Molecular structure & $9338$\\ 
62P10 & Operations research, mathematical programming & Combinatorial optimization & $8935$\\ 
35Q53 & Partial differential equations & KdV-like equations & $8528$\\ 
91B28 & Game theory, economics, social and behavioral sciences & Finance, portfolios, investment & $8366$\\ 
76D05 & Fluid mechanics & Navier-Stokes equations & $8207$\\ 
65N30 & Numerical analysis & Finite elements, Rayleigh-Ritz and Galerkin methods, finite methods & $7976$\\ 
62M10 & Statistics & Time series, auto-correlation, regression, etc. & $7926$
\end{tabular}}
\label{tab_msc}
\end{table}

\begin{table}[!h]
\centering
\caption{\small{Table of the most popular primary MSCs.}}
\vspace{5pt}
\scriptsize{
\begin{tabular}{@{}p{0.7cm}@{\hspace{2mm}}p{5.6cm}@{\hspace{2mm}}p{4.2cm}@{\hspace{2mm}}r@{}}
\textbf{MSC} & & & \textbf{No. of}\\
\textbf{code} & \textbf{$2$-char MSC name} & \textbf{MSC name} & \textbf{works}\\
\hline
74S05 & Mechanics of deformable solids & Finite element methods & $7620$\\ 
01A70 & History and biography & Biographies, obituaries, personalia, bibliographies & $5983$\\ 
68T05 & Computer science & Learning and adaptive systems & $5943$\\ 
90B35 & Operations research, mathematical programming & Scheduling theory, deterministic & $5707$\\ 
91B28 & Game theory, economics, social and behavioral sciences & Finance, portfolios, investment & $5386$\\ 
62P10 & Operations research, mathematical programming & Combinatorial optimization & $5316$\\
68U99 & Computer science & None of the above, but in this section & $5193$\\ 
62-99 & Statistics & Other applications & $5073$\\
35Q53 & Partial differential equations & KdV-like equations & $4792$\\ 
90B30 & Operations research, mathematical programming & Production models & $4656$\\ 
\end{tabular}}
\label{tab_msc1}
\end{table}

\begin{figure}[!ht]
\centering
\includegraphics[width=\textwidth]{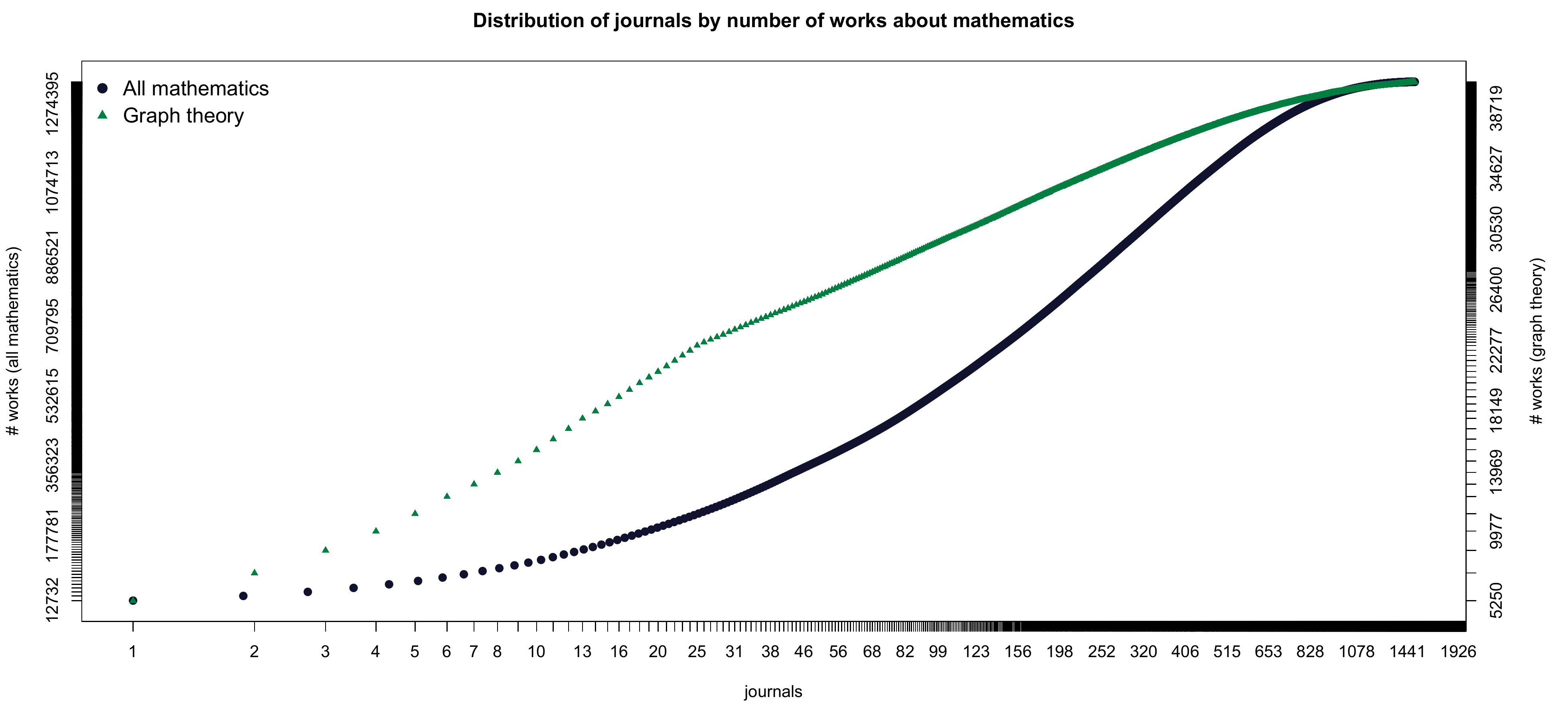}
\caption{\small{Journals in Bradford's graph form.  This figure also includes the same distribution for the field of graph theory only which is the topic of Section \ref{05cxx}.}}
\label{distj}
\end{figure}

The sequence of journals from the time-period of 1990-2010 is shown in Fig.~\ref{distj} in a shape of Bradford's graph. 
Values on the x-axis are shown in a logarithmic scale. Each dot represents one journal. Values the on y-axis are cumulative sums of indexed works in journals. Journals on the left contain the largest number of indexed works. These journals are:
\begin{itemize}
\item $32132$ works -- Journal of Physics A: Mathematical and General
\item $25464$ works -- Journal of Mathematical Analysis and Applications
\item $20564$ works -- Proceedings of the American Mathematical Society
\item $20322$ works -- Applied Mathematics and Computation
\item $18110$ works -- European Journal of Operational Research
\end{itemize}
 Journals in the right corner have published just two works indexed in ZB in a given time-period. Some of these journals are:
Journal of the History of Economic Thought, 
The Montana Mathematics Enthusiast, 
Journal of Mathematics Education, 
Vestnik Moskovskogo Universiteta. Seriya VI, 
International Journal of Energy, Environment and Economics, etc. 

%
%
\section{Collaboration Network}\label{coll}
The collaboration among mathematicians can be explored through the collaboration network. The set of nodes in the collaboration network is the set of authors and two authors are linked if they co-authored at least one work.

We determined the collaboration network as presented in \cite{collab}: $\mathbf{Co} = \A\W * \W\A$. The value of a link between two authors is equal to the number of works they have in common.
The $20$ authors with the highest numbers of co-authors are presented in Table~\ref{mcoll}. 
Since a name can belong to different authors, we checked the authors' names in the MathSciNet Authors Search \citep{ams_id}. 
The names in Table~\ref{mcoll} are divided into two columns -- the names in the first column represent a single author and the names in the second column can represent more authors. 
The third number is the number of known mathematicians with this name.

\begin{table}[!h]
\centering
\caption{\small{The authors with largest number of co-authors.}}\label{mcoll}
\scriptsize
\begin{tabular}{r|l|r||l|r|r}
&&&&& \textbf{No.~of authors}\\
&\textbf{} & \textbf{No.~of} & \textbf{} & \textbf{No.~of} & \textbf{with this}\\
\textbf{i} &\textbf{Author} & \textbf{co-authors} & \textbf{Author} & \textbf{co-authors} & \textbf{name in AMS}\\
\hline
1 & Srivastava, Hari Mohan & $347$ & Wang, Wei & $463$ & $282$\\
2 & Chen, Guanrong & $341$ &  et al. & $316$ & $$\\
3 & Alon, Noga & $288$ & Zhang, Wei & $293$ & $228$\\
4 & Pardalos, Panos M. & $212$ & Li, Wei & $277$ & $193$\\
5 & Il'in, V.A. & $195$ & Li, Jun & $244$ & $157$\\
6 & &&Wang, Hui & $232$ & $132$\\
7 & &&Wang, Yong & $224$ & $164$\\
8 & &&Wang, Jun & $223$ & $166$\\
9 & &&Zhang, Li & $218$ & $465$\\
10 & &&Li, Li & $217$ & $324$\\
11 & &&Wang, J. & $208$ & $1144$\\
12 & &&Li, Gang & $199$ & $42$\\
13 & && Zhang, Jun & $199$ & $130$\\
14 & &&Li, Ming & $193$ & $133$\\
15 & &&Wang, Y. & $192$ & $1377$
\end{tabular}
\end{table}

The subset of the most collaborative authors can be determined with $p_S$-cores \citep{cores}.
In a network $\network = (\nodes, \links, w)$ the subset $\U \subseteq \nodes$ is a $p_S$-core at level $t \in \mathbb{R}^+$ iff
\begin{itemize}
\item for each $v \in \U: p_s(v, \U) = \sum_{u \in N(v) \cap \U} w(v, u) \geq t$,
\item $\U$ is maximal.
\end{itemize}

A $p_S$-core at level $t$ in a collaboration nework is such a subnetwork in which each author's contribution to joint works with some other authors in this subnetwork is at least $t$. A lot of published works does not necessarily mean a larger collaborativeness for their author. In a computation of $p_S$-cores we are summing up the values of links. 
To neutralize the over-representation of works with many co-authors in the resulting collaboration network we used the normalized co-authorship network $\mathbf{N}$ in the computation of a collaboration network \citep{collab}: $\mathbf{N} = \diag\left(\frac{1}{\max(1,\deg(p))}\right) \cdot \W\A$. In a network $\mathbf{N}$ the values of links from a work to all of its co-authors are equal and they sum up to $1$. In \cite{collab} we calculated the normalized network $\mathbf{Ct} = \mathbf{N}^T * \mathbf{N}$ to get the contributions of authors to their works. 
Each work with $k$ authors adds to the network $\mathbf{Ct}$ a corresponding complete directed graph (with loops) on $k$ nodes. Each of its arcs has the weight $\frac{1}{k^2}$.
For the analysis of the ZB data we used a slightly changed normalized collaboration network $\mathbf{Ct}' $ which is an undirected network without loops obtained as the sum of complete undirected graphs. Each edge of a complete graph for a work with $k$ authors has the weight $\frac{2}{k\cdot (k-1)}$. The network $\mathbf{Ct}'$ can be obtained as a symmetrization of $\mathbf{N}^T * \mathbf{N}'$ and setting the diagonal values to $0$, where $\mathbf{N}' = \diag\left(\frac{1}{\max(1, \deg(w)-1)}\right) \cdot \W\A$. With this we neutralize works with many co-authors.

\begin{figure}[!ht]
\centering
\includegraphics[width=\textwidth]{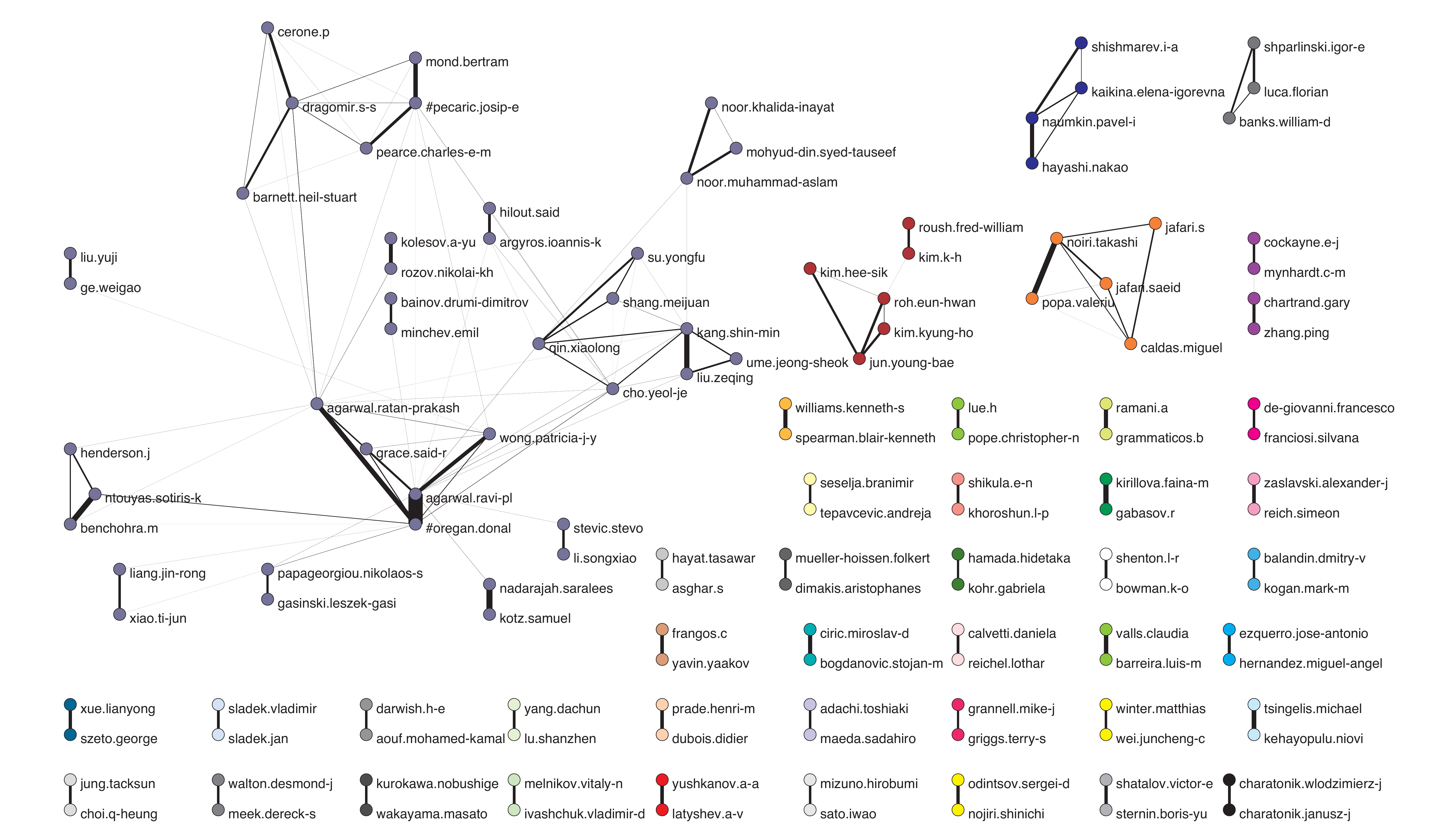}
\caption{\small{The $p_S$-core at level $t=30$ in the collaboration network $\mathbf{Ct}'$}}
\label{cores35}
\end{figure}

Fig.~\ref{cores35} shows a $p_S$-core at level $t=30$ in a normalized collaboration network from the ZB data. In the lower half we see mostly pairs of authors that represent authors that collaborate in `tandems'. Another interesting thing in this $p_S$-core is the large group of authors on the left. In this group one can notice stronger links between some authors -- darker and thicker links represent larger contribution to the works in common. Ten strongest collaboration pairs in this $p_S$-core are listed in Table~\ref{sredica10}.

\begin{table}
\caption{\small{List of the strongest collaboration pairs with the values of links between them in the third column.}}
\label{sredica10}
\begin{center}
\footnotesize
\begin{tabular}{r|llr}
\textbf{i} & \textbf{First author} & \textbf{Second author} & \textbf{Link value}  \\
\hline
  1 & Agarwal, Ravi P. 		& O'Regan, Donal 			& $49.46$ \\
  2 & Kotz, Samuel 		& Nadarajah, Saralees 			& $20.97$  \\
  3 & Ntouyas, Sotiris K. 	& Benchohra, Mouffak 			& $19.69$  \\
  4 & Popa, Valeriu 		& Noiri, Takashi 				& $19.00$  \\
  5 & Gabasov, Rafail 		& Kirillova, Faina Miha\u{i}lovna 	& $17.88$\\
  6 & Liu, Zeqing 			& Kang, Shin Min 				& $17.72$ \\
  7 & O'Regan, Donal 		& Agarwal, Ratan Prakash 		& $16.85$ \\
  8 & Pe\v{c}ari\'{c}, Josip E. 	& Mond, Bertram 			& $15.57$ \\
  9 & Kehayopulu, Niovi 	& Tsingelis, Michael 			& $15.53$  \\
 10 & Barreira, Luis M. 		& Valls, Claudia 				& $15.28$  \\
 \end{tabular}
 \end{center}
 \end{table}

The productivity of an author can be defined in different ways. Let us say, that the author is more self-sufficient if he/she has the largest value of the self-contribution to the works he/she co-authored. This information can be obtained from network $\mathbf{Cn} = \A\W * \mathbf{N}$ \citep{collab}. The link value $cn_{ij}$ in this collaboration network is equal to the contribution of the author $i$ to the works he/she wrote together with the author $j$. The weight of a loop $cn_{ii}$ is equal to the self-contribution of the author $i$ to all works that he/she co-authored and is equal to the fractional productivity defined in \citep{fractional}.

We define the self-sufficiency index $S_i$ as the proportion of author's self-contribu\-tion and the total number of his/her works. The \emph{collaborativeness} index $K_i$ is defined as complementary value to the self-sufficiency index, $K_i = 1-S_i$ \citep{collab}, that is closely related to the collaborative coefficient \citep{cc}.

The `best' mathematicians (the most productive) are listed in Table~\ref{bests} with their self-contributions denoted as $cn_{ii}$ in the second column. The total number of his/her published works is listed in the third column and the collaborativness index is listed in the fourth column. Only three names in this list can represent more than one author (as checked in the AMS Authors Search): Evans, D. J., Wang, Wei, and Zhou, Yong.

The mathematicians with the largest number of works are not necessarily on the top of the list of the "best"mathematicians. The `best'~mathematicians have a lot of works written and also a large contribution to those works. If the self-contribution of an author is almost equal to his/her total number of works, he/she tends to work alone or in small groups. The first two authors with the largest number of works are also the `best'~authors -- Edoardo Ballico and Donal O'Regan. But there is a large difference between them -- Edoardo Ballico tends to work alone and Donald O'Regan tends to work in groups. Ballico's self-contribution value is almost equal to his total number of works ($90\%$) and O'Regan's self-contribution value is equal to a half of his total number of works ($53\%$). Next in the line by the total number of works are Josip E. Pe\v cari\' c, Mohan Hari Srivastava, and Weigao Ge. The most collaborative among the authors in Table~\ref{bests} are Guanrong Chen, Ravi P. Agarwal, Lansun Chen, Jaume Llibre, and Josip E. Pečarić.

\begin{table}
\caption{\small{List of the `best'~mathematicians in 1990-2010.}}
\label{bests}
\begin{center}
\footnotesize
\begin{tabular}{r|l|rrr|}
\textbf{i} & \textbf{Author} & $\mathbf{cn_{ii}}$ & \textbf{Total} & $\mathbf{K_i}$ \\
\hline
  1 & Ballico, Edoardo        &     			$ 865.58$ & $967$  &  $0.105$ \\
  2 & O'Regan, Donal     &     				$432.80$ & $821$  &  $0.470$ \\
  3 & Argyros, Ioannis Konstantinos     &     	$353.17 $ & $373$  &  $0.053$ \\
  4 & Shelah, Saharon    &     				$ 302.49$ & $519$  &  $0.417$ \\
  5 & Verma, Ram U.     &     				$ 297.58$ & $314$  &  $0.052$ \\
  6 & Srivastava, Hari Mohan     &     		$267.17 $ & $582$  &  $0.541$ \\
  7 & Pe\v cari\' c, Josip E.     &     			$265.28 $ & $608$  &  $0.564$ \\
  8 & Papageorgiou, Nikolaos S.  &     		$263.92 $ & $418$  &  $0.369$ \\
  9 & Pachpatte, Baburao G.      &     			$ 255.00$ & $261$  &  $0.023$ \\
 10 & Maslov, Victor P.    &     			$247.08$ & $324$  &  $0.237$ \\
 11 & Agarwal, Ravi P.   &     				$244.12$ & $598$  &  $0.592$ \\
 12 & Wazwaz, Abdul-Majid     &     			$242.67 $ & $254$  &  $0.045$ \\
 13 & Noor, Muhammad Aslam      &     		$ 241.00$ & $351$  &  $0.313$ \\
 14 & Jun, Young Bae     &     				$239.40$ & $480$  &  $0.501$ \\
 15 & Dragomir, S. S.    &     				$232.73 $ & $364$  &  $0.361$ \\
 16 & Le, Maohua   &     					$ 226.17$ & $242$  &  $0.065$ \\
 17 & Ge, Weigao     &     				$225.78 $ & $504$  &  $0.552$ \\
 18 & Nadarajah, Saralees    &     			$ 213.70$ & $315$  &  $0.322$ \\
 19 & Ramm, Alexander G.    &     			$211.92 $ & $270$  &  $0.215$ \\
 20 & Stevi\' c, Stevo  &     				$ 206.20$ & $257$  &  $0.198$ \\
 21 & Gamkrelidze, R.V.     &     			$ 190.65$ & $268$  &  $0.289$ \\
 22 & Zaslavski, Alexander J.   &     			$ 187.50$ & $236$  &  $0.206$ \\
 23 & El Naschie, Mohamed Saladin  &     		$183.08$ & $186$  &  $0.016$ \\
 24 & Evans, D. J.     &     				$ 182.88$ & $356$  &  $0.486$ \\
 25 & Wang, Wei       &     				$ 178.79$ & $394$  &  $0.546$ \\
 26 & Nazarov, Serguei A.      &     			$ 176.95$ & $266$  &  $0.335$ \\
 27 & Chen, Huanyin     &     				$ 175.50$ & $206$  &  $0.148$ \\
 28 & Alzer, Horst      &     				$ 171.92 $ & $198$  &  $0.132$ \\
 29 & Luca, Florian  &     				$  167.62$ & $292$  &  $0.426$ \\
 30 & Danchev, Peter Vassilev   &     		$166.00$ & $170$  &  $0.024$ \\
 31 & Guo, Boling      &     				$ 165.08$ & $341$  &  $0.516$ \\
 32 & Nishimoto, Katsuyuki   &     			$  163.42$ & $213$  &  $0.233$ \\
 33 & Chajda, Ivan    &     				$  162.00$ & $244$  &  $0.336$ \\
 34 & Shparlinski, Igor E.        &     			$161.50$ & $292 $  &  $0.447$ \\
 35 & Owa, Shigeyoshi     &     				$  161.06$ & $331$  &  $0.513$ \\
 36 & Anastassiou, George A.      &     		$  157.08$ & $200$  &  $0.215$ \\
 37 & Noiri, Takashi     &     				$ 152.25$ & $321$  &  $0.526$ \\
 38 & Ikramov, Kh.D.      &     				$  151.08$ & $198$  &  $0.237$ \\
 39 & Jakub\'  ik, J\' an    &     				$  150.83$ & $158$  &  $0.045$ \\
 40 & Zhou, Yong      &     				$ 150.03$ & $245$  &  $0.388$ \\
 41 & Chen, Guanrong   &     				$149.54$ & $385$  &  $0.612$ \\
 42 & Biswas, Indranil      &     			$  148.83$ & $224$  &  $0.336$ \\
 43 & Llibre, Jaume      &     				$ 148.77$ & $345$  &  $0.569$ \\
 44 & Khrennikov, Andrei Yu.       &     		$148.52$ & $192$  &  $0.226$ \\
 45 & Hall, Peter G.       &     				$ 145.45$ & $294$  &  $0.505$ \\
 46 & Chen, Lansun    &     				$141.03$ & $344$  &  $0.590$ \\
 47 & Aouf, Mohamed Kamal      &     		$139.99$ & $ 251$  &  $0.442$ \\
 48 & Chen, Bang-Yen  &     				$  138.00$ & $177$  &  $0.220$ \\
 49 & Park, Sehie      &     				$ 137.25 $ & $166$  &  $0.173$ \\
 50 & Alon, Noga        &     				$ 136.37$ & $298$  &  $0.542$\\
 \hline
 \end{tabular}
 \end{center}
 \end{table}

%
%
\section{05Cxx Graph theory}\label{05cxx}
Graph theory is a subdiscipline of combinatorics 05 and its three-char MSC is $05C.$ We can look at graph theory as pure or with its applications -- MSCs from other mathematical disciplines which are by content connected to graph theory can be included.

For further analysis we took the network $\W\mathbf{M}_3$ which is a shrinken version of the network $\W\mathbf{M}$: the set of 5-char MSCs is shrinked into a set of 3-char MSCs. A combination of this network with other networks allows us to analyze the field of graph theory as it can be seen through published works.

To see which journals published the largest amount of indexed works about graph theory, we need a network $\W\mathbf{J}$ and a network $\W\mathbf{M}_3$. The values of links in the second network might be larger than one -- a work can have more MSCs with the same first 3 chars determined. We changed these values of links to 1 and multiplied networks $\W\mathbf{J}$ and $\W\mathbf{M}_3$ to get the network $\mathbf{JM}_3 = \mathbf{J}\W * b(\W\mathbf{M}_3)$, where $b(\network)$ is the binarized version of network $\network$. The link value $jm3_{jc}$ in this network is equal to the number of indexed works that were published in a journal $j$ and were classified with a classification $c$. 
We normalized this network in a similar way as we normalized the $\W\A$ network to get the normalized collaboration network: $n(\mathbf{JM}_3) = \diag\left(\frac{1}{\textrm{weighted deg} (j)}\right) \mathbf{JM}_3$. 
The weighted degree of a node is equal to the sum of incident links values. 
The sum of incident links' values of each journal in the network $n(\mathbf{JM}_3)$ is now equal to 1. 

We took a look at link values from journals to the graph theory classifications. The link values represent the percentages of indexed works in the ZB published in the selected journal that are classified with a graph theory MSC. 
In the left column of the Table~\ref{gt_journals} are listed the journals with the largest percentages of such works. In the right column of the Table~\ref{gt_journals} are listed the journals that have largest percentages of indexed works in the ZB about graph theory with its applications included.  MSCs that represent graph theory's applications are 68R10, 81Q30, 81T15, 82B20, 82C20, 90C35, 92E10, 94C15, 05E30, 57M15, 57M25, 20F65, 90B10, 05B30, 05D10, 91A43, 91A46, 60B20, 91D30, 68R10, 68W05, 81Q30, 81T15, 82B20, 82C20, 90C35, 92E10, 94C15, and all that start with 90B.

The difference in both lists is easily seen. There is one journal (The European Physical Journal B. Condensed Matter) from which only works with at least one classification from graph theory or its applications were included in the ZB.

\begin{table}[!ht]
\centering
\caption{\small{Journals with the largest percentages of indexed works about graph theory in the time-period 1990-2010: pure graph theory (left), graph theory with applications (right).}}
\vspace{5pt}
\scriptsize{
\begin{tabular}{p{3.3cm}r||p{4.3cm}r}
\multicolumn{2}{l}{\textbf{Pure graph theory}}& \multicolumn{2}{l}{\textbf{Graph theory and its applications}}\\
\hline
Journal of Graph Theory (0364-9024, 1097-0118) & $89.15\%$ & The European Physical Journal B. Condensed Matter (1434-6028) & $100.00\%$\\
AKCE International Journal of Graphs and Combinatorics (0972-8600) & $82.55\%$ & Journal of Graph Theory (0364-9024, 1097-0118) & $95.87\%$\\
Journal of Combinatorial Theory. Series B (0095-8956) & $68.85\%$ & AKCE International Journal of Graphs and Combinatorics (0972-8600) & $88.89\%$\\
Graphs and Combinatorics (0911-0119, 1435-5914) & $59.45\%$ & Journal of Combinatorial Theory. Series B (0095-8956) & $83.20\%$\\
Ars Combinatoria (0381-7032) & $50.04\%$ & ITS Journal (1024-8072) & $80.77\%$\\
The Australasian Journal of Combinatorics (1034-4942) & $46.96\%$ &  International Journal of Flexible Manufacturing Systems (0920-6299, 1572-9370) & $79.44\%$\\
JCMCC. The Journal of Combinatorial Mathematics and Combinatorial Computing    (0835-3026) & $43.81\%$ & Graphs and Combinatorics (0911-0119, 1435-5914) & $76.49\%$\\
Ars Mathematica Contemporanea (1855-3966, 1855-3974) & $42.86\%$ & International Journal of Production Research    (0020-7543, 1366-588X) & $76.48\%$\\
Congressus Numerantium (0384-9864) & $42.81\%$ & Match (0340-6253) & $75.05\%$\\
Match (0340-6253) & $42.80\%$ & Journal of Graph Algorithms and Applications (1526-1719) & $74.72\%$\\
Discrete Mathematics (0012-365X) & $42.39\%$ & Location Science (0966-8349) & $74.29\%$\\
Bulletin of the Institute of Combinatorics and its Applications (1183-1278) & $42.28\%$ &  Journal of Scheduling (1094-6136, 1099-1425) & $68.84\%$\\
Advances and Applications in Discrete Mathematics (0974-1658) & $42.27\%$ & Journal of Interconnection Networks (0219-2659) & $66.67\%$\\
Combinatorica (0209-9683) & $33.99\%$ & Studies in Locational Analysis (1105-5162) & $66.48\%$\\
Combinatorics, Probability and Computing (0963-5483, 1469-2163) & $33.76\%$ & Networks (0028-3045, 1097-0037) & $66.04\%$\\
College Mathematics Journal (0746-8342) & $33.33\%$ & Transportation Science (0041-1655) & $63.77\%$\\
International Journal of Mathematical Combinatorics (1937-1055) & $29.29\%$ &  Networks and Spatial Economics (1566-113X, 1572-9427) & $62.00\%$\\
Random Structures \& Algorithms (1042-9832, 1098-2418) & $29.06\%$ & The Australasian Journal of Combinatorics (1034-4942) & $60.31\%$\\
Discussiones Mathematicae. Graph Theory (1234-3099) & $28.95\%$ & Ars Combinatoria (0381-7032) & $59.96\%$\\
Journal of Combinatorics, Information \& System Sciences (0250-9628) & $28.24\%$ & JCMCC. The Journal of Combinatorial Mathematics and Combinatorial Computing (0835-3026) & $59.87\%$\\
\end{tabular}}
\label{gt_journals}
\end{table}

Another way of determining journals that published a lot of works about graph theory is using biases~\citep{grcar}. The bias of a journal for or against any branch of mathematics is
$$
\textrm{bias} = \log_2 \frac{\textrm{fraction of works about the subject in the journal}}{\textrm{fraction of works about the subject in all of mathematics}}.
$$ 

This value basically tells us if some journal is favoring a selected branch or subject of mathematics (positive value) or if it is hindering it (negative value). If the value of bias is equal to zero, the journal published relatively as many works about the selected branch or subject of mathematics as all journals together did.

In Table \ref{gt_bias_p} the journals with the largest positive biases for the graph theory are listed, and in Table \ref{gt_bias_n} are the journals with the largest negative biases for the graph theory. We include in the calculation of the bias value only the journals that published at least $50$ works  indexed in the ZB database. An author can use the bias value for his/her topic to determine the best journals for submitting his/her work. A positive bias of a journal for the  selected topic means that this journal is more likely to publish a work about this topic; and a negative bias of a journal for a topic means that this journal is more likely to reject a work about this topic.

\begin{table}[!ht]
\centering
\caption{\small{Journals with the largest positive biases for graph theory in the time period 1990-2010.}}
\vspace{5pt}
\scriptsize{
\begin{tabular}{lr}
\textbf{Journal} & \textbf{Bias}\\
\hline
Journal of Graph Theory (0364-9024) & $6.035$\\
Discussiones Mathematicae. Graph Theory (1234-3099) & $6.023$\\
AKCE International Journal of Graphs and Combinatorics (0972-8600) & $5.928$\\
Journal of Combinatorial Theory. Series B (0095-8956) & $5.774$\\
Graphs and Combinatorics (0911-0119) & $5.618$\\
Applicable Analysis and Discrete Mathematics (1452-8630) & $5.604$\\
Ars Combinatoria. The Canadian Journal of Combinatorics (0381-7032) & $5.297$\\
The Australasian Journal of Combinatorics (1034-4942) & $5.265$\\
JCMCC. The Journal of Combinatorial Mathematics and Combinatorial Computing (1983-0823) & $5.238$\\
MATCH - Communications in Mathematical and in Computer Chemistry (0340-6253) & $5.233$\\
\end{tabular}}
\label{gt_bias_p}
\end{table}

\begin{table}[!ht]
\centering
\caption{\small{Journals with the largest negative biases for graph theory in the time period 1990-2010.}}
\vspace{5pt}
\scriptsize{
\begin{tabular}{p{10cm}r}
\textbf{Journal} & \textbf{Bias}\\
\hline
International Journal of Solids and Structures (0020-7683) & $-7.765$\\
Journal of Differential Equations (0022-0396) & $-7.427$\\
International Journal of Modern Physics A. Particles and Fields, Gravitation and Cosmology (0217-751X) & $-7.127$\\
Classical and Quantum Gravity. An International Journal of Gravitational Physics, Cosmology, Geometry and Field Theory (0264-9381) & $-6.982$\\
Modern Physics Letters A. Particles and Fields, Gravitation, Cosmology, Nuclear Physics (0217-7323) & $-6.797$\\
Systems \& Control Letters (0167-6911) & $-6.516$\\
Applicable Analysis. An International Journal (0003-6811) & $-6.472$\\
Acta Arithmetica (0065-1036) & $-6.357$\\
Nonlinear Analysis. Theory, Methods \& Applications. Series A: Theory and Methods. An International Multidisciplinary Journal (0362-546X) & $-6.124$\\
Annals of Physics (0003-4916) & $-6.070$\\
\end{tabular}}
\label{gt_bias_n}
\end{table}

For further analysis we used the network $\W\mathbf{M}^{[05C]},$ which is a network $\W\mathbf{M}$ with the second set of nodes restricted so that only MSCs from graph theory remain. In order to get the network $\W\mathbf{M}^{[05C]}$, we first made a partition of classifications $\sigma$ in which all 05C classifications are in one class and the other classifications are in another class. With the partition $\sigma$ we extracted the subnetwork $\W\mathbf{M}_{\sigma}$ from the network $\W\mathbf{M}$. The network $\W\mathbf{M}_{\sigma}$ contains all works and only 05C classifications. Then we determined the outdegree partition of works $\tau$ and removed from the network $\W\mathbf{M}_{\sigma}$ all nodes (works) with outdegree $0$. The works with outdegree greater than $0$ have at least one MSC from graph theory. The resulting network is $\W\mathbf{M}^{[05C]}.$

We used the partition of works $\tau$ on networks $\W\A$, $\W\mathbf{J}$, and $\W\mathbf{K}$ to extract networks $\W\A^{[05C]}$, $\W\mathbf{J}^{[05C]}$, and $\W\mathbf{K}^{[05C]},$ respectively, in which are included only works about graph theory and their authors, journals in which they were published, and used keywords.

We examined degrees of nodes in the obtained networks to determine distributions of different data as we did for the whole set of works in Section \ref{distributions}.

The distribution of works about graph theory by the number of authors is presented in Fig.~\ref{density_gt} in Section \ref{distributions} in the top diagram. The distribution is shown in the same diagram as the distribution of all analyzed works by the number of authors. Both distributions are similar. More than $\frac{1}{3}$ of all works ($34.34\%$) were written by a single author and even more ($36.86\%$) by a pair of authors. $19$ works do not have any author determined.

The distribution of works about graph theory by the number of keywords is shown in Fig.~\ref{density_gt} in the middle. This distribution has a higher peak at a lower value ($7$) than the distribution of all analyzed works by the number of assigned keywords. Approximately $65 \%$ of all works have the number of keywords between $5$ and $10$. 

The distribution of works about graph theory by the number of MSCs is shown in Fig.~\ref{density_gt} at the bottom figure. This distribution is also almost the same as the distribution of all analyzed works by the number of MSCs. Approximately one third of all works ($30.27\%$) were classified with two MSCs and $47.38\%$ of all works were classified with one or three MSCs.

The distribution of authors by the number of works about graph theory they co-authored is displayed in Fig.~\ref{dist}  in Section \ref{distributions} in the top figure together with the distribution of authors by the number of all works they co-authored. For example, the lighter dot in the upper left corner represents $13801$ authors that in the time-period 1990-2010 each co-authored only one work. Both distributions looks alike.

The distribution of keywords by the number of works about graph theory they describe is shown in Fig.~\ref{dist} in the second figure together with the distribution of keywords by the number of all works they describe. The lighter dot in the upper left corner represents $6231$ keywords that each was used in description of only one work in the time-period of 1990-2010. Again, the shape of the distribution is typical for the power law $f_n = cn^{-\alpha}$ for $\alpha = 1.72$, which is a bit smaller than the value of $\alpha$ for the distribution of keywords according to all works ($\alpha = 1.85$).

And finally, the distribution of MSCs by the number of works about graph theory that were classified with them is shown in Fig.~\ref{dist} at the bottom figure together with the distribution of MSCs by the number of all analyzed works that were classified with given MSCs. The lighter dot in the upper left corner represents $1336$ MSCs, that each was included in the classification of only one work about graph theory in the time-period of 1990-2010. This distribution has a higher value at the beginning (at value $1$) and drops faster than the distribution of MSCs according to all analyzed works.

The sequence of journals sorted in a decreasing order by the number of indexed works about graph theory in the time-period 1990-2010 is shown in Fig.~\ref{distj} together with the sequence of journals in a decreasing order by the number of all indexed works in the time period 1990-2010.
Journals on the left have published the largest numbers of works.
The Bradford's graph form of the sequence of journals about graph theory coincide with the Bradford's graph form of the sequnce of journals about all mathematics only in the end -- on the right side.

We used the partition of works $\tau$ on the network $\W\A$ to extract the network $\W\A^{[05C]}$ in which are included only the works about graph theory and their authors. With the input degree partition of the second set of nodes in the network $\W\A^{[05C]}$ we got the list of authors that published largest amounts of works about graph theory. Another way to see which authors published largest amounts of works about graph theory is to look at the values on the loops in the normalized collaboration network (Table~\ref{gt_nauthors}). We checked the uniqueness of names in this list with the AMS Authors Search and only two names possibly represent more than one author: Liu, Guizhen (two authors) and Zhang, Ping ($58$ authors).

\begin{table}[!ht]
\centering
\caption{\small{The list of $20$ authors with the largest contributions to their works about graph theory in the time-period 1990-2010.}}
\vspace{5pt}
\footnotesize{
\begin{tabular}{r|l|rrr}
\textbf{i} & \textbf{Author} & $\mathbf{cn_{ii}}$ & \textbf{Total} & $\mathbf{K_i}$ \\
\hline 
1 & Volkmann, Lutz & 			$123.55$ & 	$216$ &	$0.428 $\\
2 & Henning, Michael A. & 		$110.87$ & 	$232$ &$0.522 $\\
3 & Liu, Yanpei & 			$102.42$ & 	$196$ &	$0.478 $\\
4 & Alon, Noga & 				$85.39$ & 	$177$ & 	$0.518 $\\
5 & Tuza, Zsolt & 			$77.05$ & 	$150$ &	$0.486 $\\
6 & Zhu, Xuding & 			$76.85$ & 	$132$ & 	$0.418 $\\
7 & Gutman, Ivan & 			$70.13$ & 	$143$ & 	$0.510 $\\
8 & Thomassen, Carsten &		$68.83$ & 	$82$ & 	$0.161 $\\
9 & Mohar, Bojan & 			$67.85$ & 	$111$ & 	$0.389 $\\
10 & Liu, Guizhen & 			$67.28$ & 	$137$ & 	$0.509 $\\
11 & Liu, Bolian & 			$63.67$ & 	$119$ &	$0.465 $\\
12 & Klav\v{z}ar, Sandi & 		$62.74$ & 	$129$ & 	$0.514 $\\
13 & Bollob\'as, B\'ela & 		$62.53$ & 	$143$ & 	$0.563 $\\
14 & Zhang, Ping & 			$60.47$ & 	$157$ & 	$0.615 $\\ 
15 & Li, Xueliang & 			$60.40$ & 	$136$ & 	$0.556 $\\
16 & R\"odl, Vojt\v{e}ch & 		$57.98$ & 	$146$ & 	$0.603 $\\
17 & Zhang, Zhongfu & 			$57.07$ & 	$162$ &	$0.648 $\\
18 & McKee, Terry A. & 		$54.98$ & 	$64$ & 	$0.141 $\\
19 & Zelinka, Bohdan & 		$54.50$ & 	$57$ & 	$0.044 $\\
20 & Yuster, Raphael & 			$52.56$ & 	$79$ & 	$0.335 $\\
\end{tabular}}
\label{gt_nauthors}
\end{table}

\begin{figure}[!h]
\centering
\includegraphics[width=\textwidth]{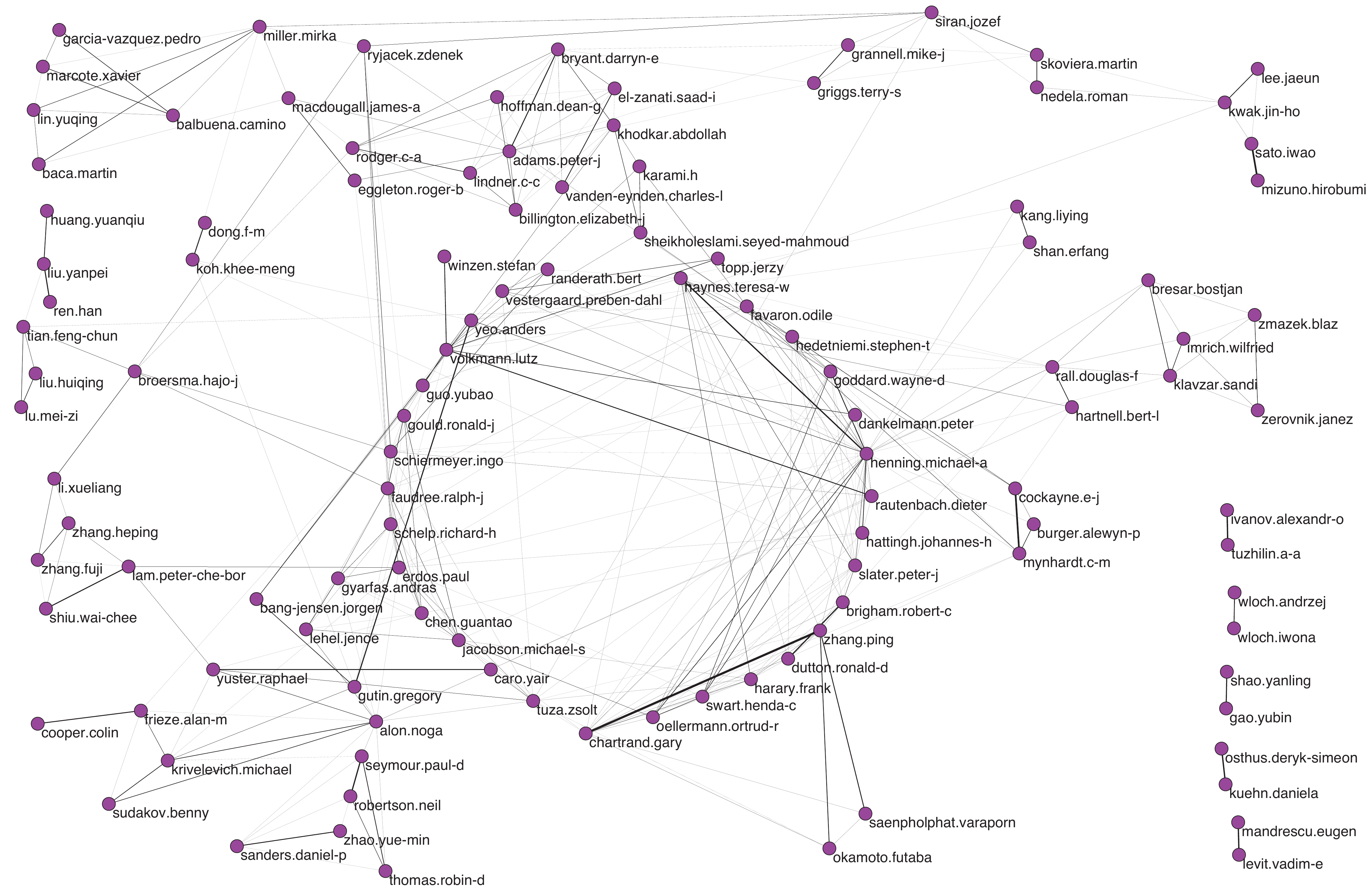}
\caption{\small{The $p_S$-core at level $t=3.50$ in the collaboration network $\mathbf{Ct}$ of graph theorists.}}
\label{cores05C}
\end{figure}

As we searched for the strongest collaboration ties in the collaboration network among all mathematicians, we did the same for graph theorists. We determined the normalized collaboration network $\mathbf{Ct}^{[05C]}$ for graph theorists -- using the normalized $\W\A^{[05C]}$ network. The $p_S$-core at level $t=3.5$ is presented in Fig.~\ref{cores05C}. There are only few pairs of collaborators and one big group. One can notice stronger collaborations (darker and thicker links) inside subgroups of this group and these subgroups are linked to each other with weaker collaborations (lighter links).

\begin{figure}[!h]
\centering
\includegraphics[width=0.80\textwidth]{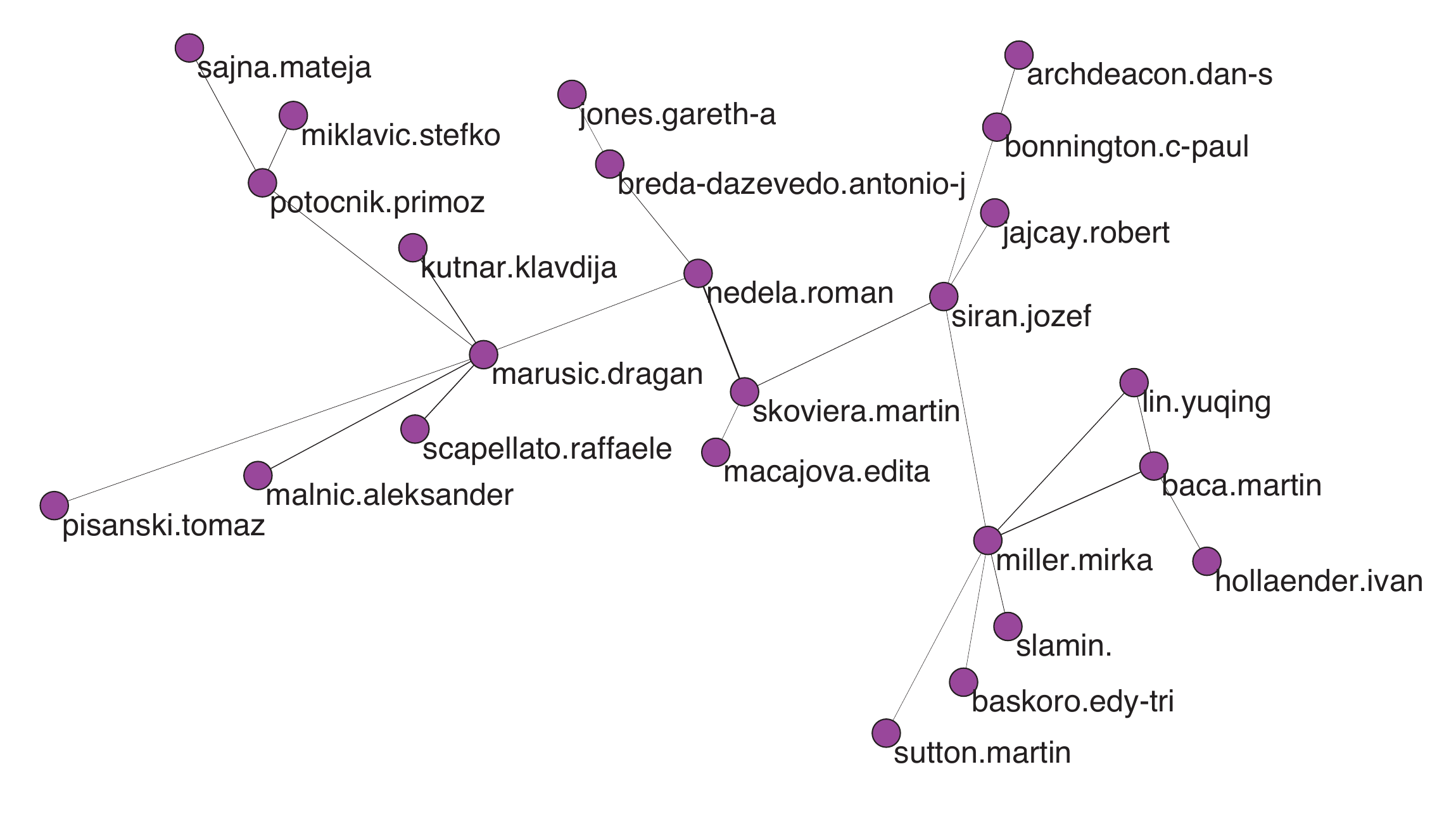}
\caption{\small{A link island of graph theorists in the normalized collaboration network $\mathbf{Ct}$ with a subgroup of Slovenian and Slovak graph theorists.}}
\label{island1}
\end{figure}

\begin{figure}[!h]
\centering
\includegraphics[width=0.95\textwidth]{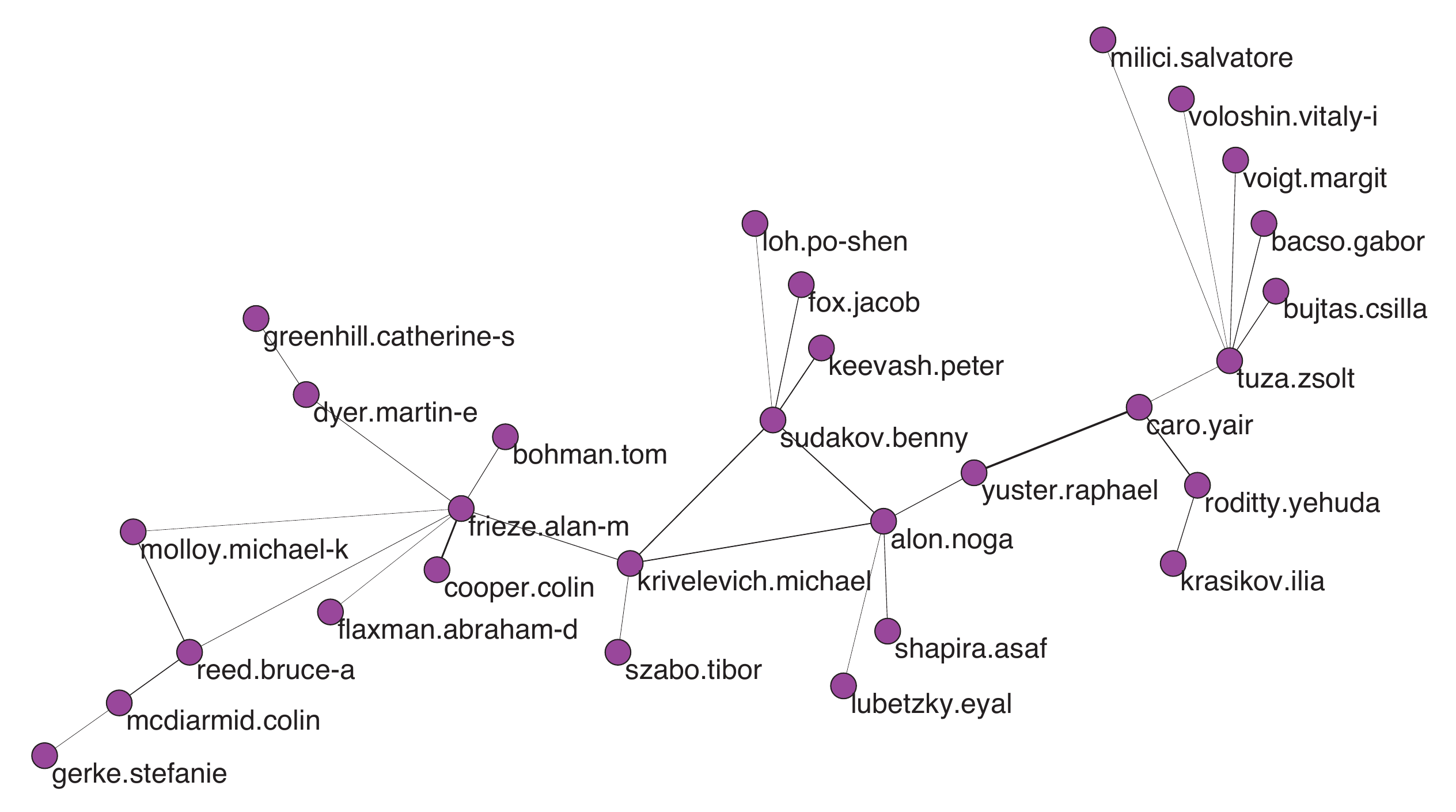}
\caption{\small{A link island of graph theorists in the normalized collaboration network $\mathbf{Ct}$ with Noga Alon in the middle.}}
\label{island2}
\end{figure}

\begin{figure}[!h]
\centering
\includegraphics[width=0.90\textwidth]{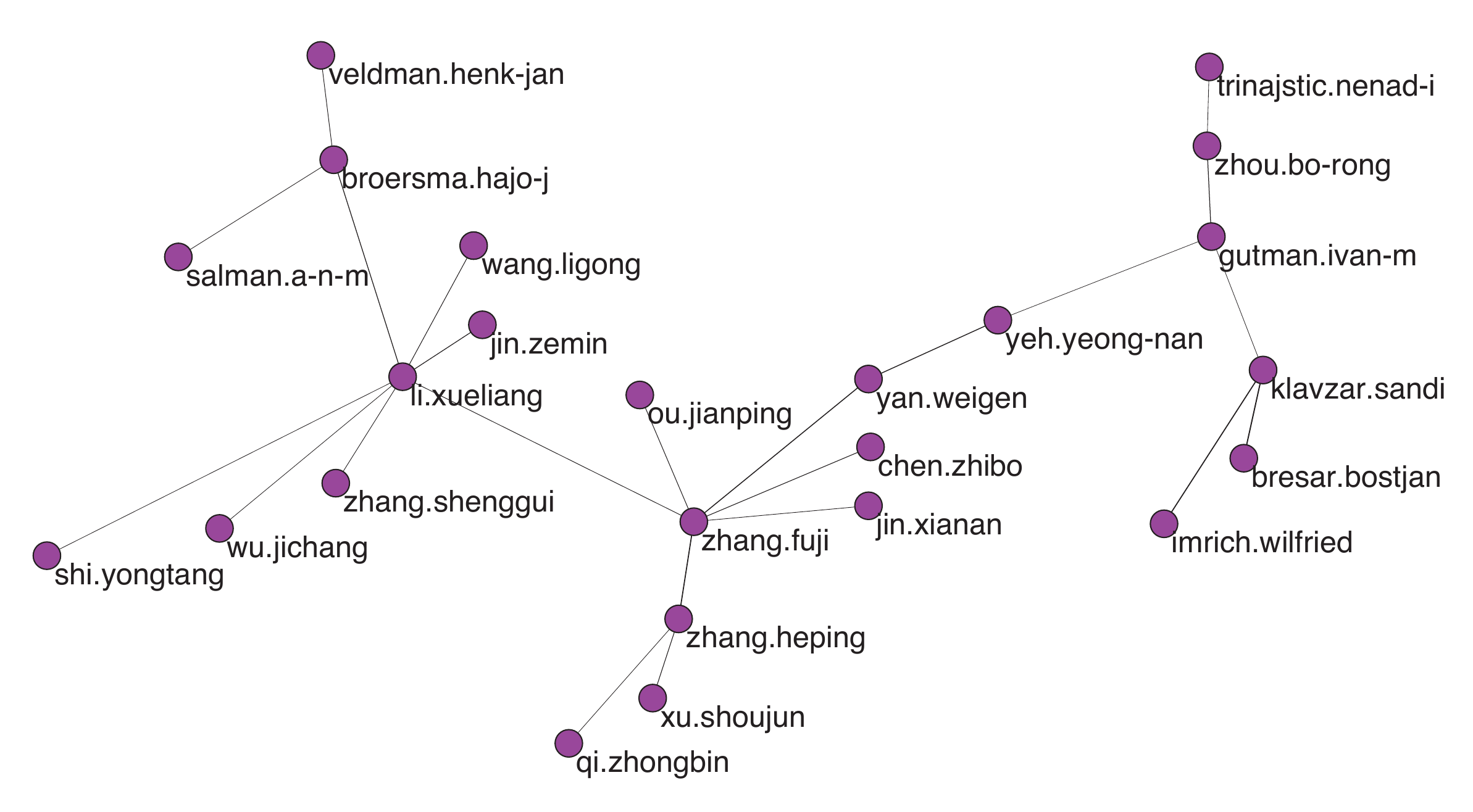}
\caption{\small{A link island of graph theorists in the normalized collaboration network $\mathbf{Ct}$ with mostly Asian names.}}
\label{island4}
\end{figure}

Another way of identifying strong collaboration groups among graph theorists is using link islands. A link island in a network $\network = (\nodes, \n{E}, w)$ is a subnetwork $\n{M} = (\n{U}, \n{F}, w)$ such that there exists a spanning tree $\mathcal{T}$, such that the values of links with exactly one end node in $\mathcal{U}$ are smaller or equal to the smallest value of links of the tree $\mathcal{T}$.
The link islands determine the locally important subnetworks.
In Fig.~\ref{island1}, \ref{island2}, \ref{island4} three link islands of the size between $10$ and $30$ for the graph theorists in the normalized collaboration network $\mathbf{Ct}$ are presented. For details see the slides
Zaver\v snik, M., \& Batagelj, V. (2004): 
\emph{Islands} that were presented on the XXIV. International Sunbelt Social Network Conference in 
  Portoro\v{z}, Slovenia, available at http://vlado.fmf.uni-lj.si/pub/networks/doc/sunbelt/islands.pdf.

Many works about graph theory have also some other classifications besides graph theory. Multiple classifications for a work are representing the interdisciplinarity of a work. We used the network $\W\mathbf{M}^{[05C]}$ to get the list of classification that coappeared within works about graph theory the most. To get this information, we used a partition of works $\tau$ in the network $\W\mathbf{M}$ to get a subnetwork of works about graph theory and all classifications $\W_{\tau}\mathbf{M}$. We shrinked the set of works into a single node. Classifications with the largest  weighted input degrees represent mathematical areas that work interdisciplinary with the graph theory the most. These classifications are listed in Table~\ref{gt_mscs}. Each classification is defined with the MSC code in the first column, its name in the third column and a 2-char classification name (mathematical discipline). Classifications are arranged according to the value in the last column -- the number of works about the graph theory that were classified also with those classifications.

\begin{table}[!ht]
\centering
\caption{\small{A list of MSCs which coappeared most frequently with the graph theory classifications in the time-period 1990-2010.}}
\vspace{5pt}
\scriptsize{
\begin{tabular}{p{0.7cm}p{4.6cm}p{4.3cm}p{0.7cm}}
\textbf{MSC code} & \textbf{$2$-char MSC name} & \textbf{MSC name} & \textbf{No. of works}\\
\hline
68R10 & Computer science & Graph Theory & $3528$\\ 
68Q25 & Computer science & Analysis of algorithms and problem complexity & $1140$\\ 
90C35 & Operations research, mathematical programming & Programming involving graphs or networks & $868$\\ 
92E10 & Biology and other natural sciences & Molecular structure & $591$\\ 
90C27 & Operations research, mathematical programming & Combinatorial optimization & $572$\\ 
60C05 & Probability theory and stochastic processes & Combinatorial probability & $521$\\ 
05A15 & Combinatorics & Exact enumeration problems, generating functions & $492$\\ 
15A18 & Linear and multilinear algebra; matrix theory & Eigenvalues, singular values, and eigenvectors & $392$\\ 
57M15 & Manifolds and cell complexes & Relations with graph theory & $381$\\ 
05B35 & Combinatorics & Matroids, geometric lattices & $334$\\ 
94C15 & Information and communication, circuits & Applications of graph theory & $317$\\ 
68W25 & Computer science & Approximation algorithms & $315$\\ 
05E30 & Combinatorics & Association schemes, strongly regular graphs & $312$\\ 
06A07 & Order, lattices, ordered algebraic structures & Combinatorics of partially ordered sets & $291$\\ 
90B10 & Operations research, mathematical programming & Network models, deterministic & $283$\\ 
20B25 & Group theory and generalizations & Finite automorphism groups of algebraic, geometric, or combinatorial structures & $280$\\ 
20D60 & Group theory and generalizations & Arithmetic and combinatorial problems & $275$\\ 
68M10 & Computer science & Network design and communication & $274$\\ 
91A43 & Game theory, economics, social and behavioral sciences & Games involving graphs & $237$\\ 
05B20 & Combinatorics & Matrices & $222$\\
\end{tabular}}
\label{gt_mscs}
\end{table}

Records for most of the works contain information about keywords. Some keywords are common in all areas of mathematics and some are used only in few areas. With a right weightening of keywords we can sort them by their importance for different areas of mathematics. We used the TF-IDF weightening \citep{tf-idf-clanek}. Areas of mathematics can be determined by MSCs. We multiplied networks $\mathbf{M}\W$ and $\W\mathbf{K}$ in order to obtain the network $\mathbf{MK}$. MSCs were shrinked according to $2$-char MSC codes.

All keywords, used in all areas get the value zero in TF-IDF weightening. Others get values:
$$
\textrm{TF-IDF}(keyword, MSC) = \textrm{TF}(keyword, MSC) \times \textrm{IDF}(keyword)
$$
$$
\textrm{TF}(keyword, MSC) = \frac{\textrm{Value on the link between }keyword\textrm{ and }MSC}{\textrm{Sum of values of all links from }MSC}
$$
$$
\textrm{IDF}(keyword) = \log \frac{\textrm{No. of MSCs}}{\textrm{No. of MSCs linked to }keyword}
$$

There are $12460$ keywords with a non-zero TF-IDF value linked to the MSCs of graph theory. The largest values have the keywords listed in Table~\ref{ti_gt}. In the table are also listed absolute frequencies of keywords within graph theory and within all mathematics.

\begin{table}[!ht]
\centering
\caption{\small{A list of keywords with the highest TF-IDF value for the $3$-char MSC $05C$ in the time-period 1990-2010.}}
\vspace{5pt}
\footnotesize{
\begin{tabular}{lrrr}
\textbf{Keyword} & \textbf{No. of appearances} & \textbf{No. of all} & \textbf{TF-IDF value}\\
 &\textbf{within graph theory}& \textbf{appearances} & \textbf{($\cdot 10^{-4}$)}\\
\hline
Coloring & $4133$ & $6676$ & $45.24$\\
Digraph & $4049$ & $5695$ & $44.32$\\
Chromatic & $3958$ & $5138$ & $43.32$\\
Subgraph & $2298$ & $3473$ & $35.31$\\
Domination & $2500$ & $3883$ & $27.36$\\
Clique & $1788$ & $3169$ & $25.29$\\
Vertex & $4611$ & $12981$ & $25.23$\\
Hypergraph & $2258$ & $3808$ & $24.71$\\
Bipartite & $2776$ & $5045$ & $24.08$\\
Tournament & $1335$ & $2341$ & $21.92$\\
Matching & $1038$ & $1891$ & $17.04$\\
Label & $1960$ & $4656$ & $17.00$\\
Ramsey & $1305$ & $3148$ & $16.58$\\
Claw & $625$ & $715$ & $15.02$\\
Colour & $1278$ & $2385$ & $13.99$\\
Girth & $736$ & $954$ & $13.93$\\
Connectivity & $2528$ & $5462$ & $13.83$\\
Hamiltonicity & $465$ & $551$ & $11.96$\\
Match & $2159$ & $15140$ & $11.82$\\
Chordal & $738$ & $1384$ & $11.34$
\end{tabular}}
\label{ti_gt}
\end{table}

\section{Conclusions}
The bibliographic data can be analyzed in many ways. In this paper we present some network analysis approaches applied to the Zentralblatt MATH database that stores information about mathematical publications. Through the results of our analysis of the ZB data from a time period 1990-2010 we conclude that mathematicians tend to work alone or in small groups. They also work in a specific area of mathematics. This can be seen from the small number of MSCs that classified each work and the small number of keywords per work.

Because the data entries in the database are only partially standardized there are some problems with the data. These problems can cause irregularities in the results. We solved some of the problems (for example the unification of journals) and partially solved some other problems (for example the unification of the names of authors).

We took a closer look at works about graph theory and determined journals that are `friendly' to graph theory, the best graph theorists according to their contribution to the works they co-authored, other areas of mathematics that are closely connected to the graph theory through publications, and the keywords characteristic for the graph theory.

The network multiplication of compatible two-mode networks allows us to compute different derived networks. A network $\mathbf{AJ} = \mathbf{A}\W * \W\mathbf{J}$ stores the information of the number of indexed works that were written by some author and published in some journal. This network can be analyzed or used further to produce new networks. One possibility is to multiply it by its transpose and obtain the network $\mathbf{JJ} = b(\mathbf{JA} * \mathbf{AJ})$. Two journals in this network are linked if there exists an author that published at least one indexed work in both journals.
Another possibility is to use binarized networks: $\mathbf{JJ}_{\A} = b(\mathbf{JA}) * b(\mathbf{AJ})$. In it, the weight of a link between two journals is equal to the number of authors that published in both journals. Using approaches presented in this paper, we could analyze similarities among indexed journals.

This is just an example of what could be done in the network analysis of the ZB data in the future. In our analysis we did not consider the information about the publication year. We plan to do the temporal analysis of the ZB data and to present the results in another paper.

\begin{acknowledgements}
We thank prof. Bernd Wegner and his associates at FIZ Karlsruhe for providing the
data, and prof.~Toma\v z Pisanski and dr.~Boris Horvat for their joint part of the work on this project. We also thank Selena Praprotnik and anonymous referees for checking the text and suggesting several improvements. 

The first author was financed in part by the European Union, European Social Fund. The work was supported in part by the ARRS, Slovenia, grant J5-5537, as well as by a grant within the EUROCORES Programme EUROGIGA (project GReGAS) of the European Science Foundation. 
\end{acknowledgements}


\begin{thebibliography}{00}
\bibitem[Ajiferuke, Burell and Tague(1988)]{cc}
Ajiferuke, I., Burell, Q., Tague, J. (1988).
Collaborative coefficient: A single measure of the degree of collaboration in research.
\emph{Scientometrics, 14(5--6),} 421-433.doi: 10.1007/BF02017100

\bibitem[Batagelj and Cerin\v sek(2013)]{collab}
Batagelj, V., \& Cerin\v sek, M. (2013).
On bibliographic networks.
\emph{Scientometrics, 96(3),} 845--864. doi: 10.1007/s11192-012-0940-1

\bibitem[Batagelj, Doreian, Ferligoj and Kej\v zar(2014)]{tempo}
Batagelj, V., Doreian, P., Ferligoj, A., Kej\v zar, N. (2014).
\emph{Understanding Large Temporal Networks and Spatial Networks:
Exploration, Pattern Searching, Visualization and Network Evolution.}
New York: Wiley.

\bibitem[Batagelj and Mrvar(2014)]{pajek}
Batagelj, V.  \& Mrvar, A. (2014).
Pajek and Pajek-XXL -- Program for analysis and visualization of large networks.
http://mrvar.fdv.uni-lj.si/pajek/pajekman.pdf.
Accessed 7 May 2014.

\bibitem[Batagelj and Zaver\v snik(2011)]{cores}
Batagelj, V., \& Zaver\v snik, M. (2011).
Fast algorithms for determining (generalized) core groups in social networks.
\emph{Advances in Data Analysis and Classification, 5(2),} 129--145. doi: 10.1007/s11634-010-0079-y

\bibitem[Clauset, Shalizim and Newman(2009)]{powerlaw}
Clauset, A., Shalizi, C. R., Newman, M. E. J. (2009).
Power-law distributions in empirical data.
\emph{SIAM Review, 51(4),} 661--703. doi: 10.1137/070710111

\bibitem[De Nooy, Mrvar and Batagelj(2012)]{pajek1}
De Nooy, W., Mrvar, A., \& Batagelj, V. (2012).
\emph{Exploratory Social Network Analysis with Pajek (Structural Analysis in the Social Sciences);
Revised and Expanded Second Edition.}
Cambridge; New York: Cambridge University Press.

\bibitem[Garfield(1979)]{garfield}
Garfield, E. (1979).
\emph{Citation Indexing: Its Theory and Application in Science, Technology, and Humanities.}
New York: Wiley.

\bibitem[Grcar(2010)]{grcar}
Grcar, J. F. (2010).
Topical bias in generalist mathematics journals.
\emph{Notices of the AMS, 57(11),} 1421--1424.

\bibitem[Price and Beaver(1966)]{fractional}
De Solla Price, D., Beaver, D. de B. (1966).
Collaboration in an invisible college.
\emph{American Psychologist, 21(11),} 1011--1018.

\bibitem[Robertson(2004)]{tf-idf-clanek}
Robertson, S. (2004).
Understanding Inverse Document Frequency: On theoretical arguments for IDF.
\emph{Journal of Documentation, 60(5),} 503--520. doi: 10.1108/00220410410560582

\bibitem[TePaske-King and Richert(2001)]{ams_id}
TePaske-King, P.  \& Richert, N. (2001).
Database Reviews and Reports. The Identification of Authors in the Mathematical Reviews Database.
http://www.istl.org/01-summer/databases.html.
Accessed 7 May 2014.

\bibitem[Wegner and Werner(2010)]{mscji}
Wegner, B.  \& Werner, D. (2010).
Zentralblatt MATH, Mathematics Subject Classification 2010.
http://www.mathem.pub.ro/dept/MSC-2010\_ZBL.pdf.
Accessed 7 May 2014.

\end{thebibliography}

\end{document}